\documentclass[%
 reprint,
 superscriptaddress,
 amsmath,amssymb,
 aps,
 prx
]{revtex4-1}

\usepackage{braket}
\usepackage{graphicx}
\usepackage{dcolumn}
\usepackage{bm}
\usepackage{siunitx}
\usepackage{xcolor} 
\usepackage{tabularx}
\usepackage{dsfont}					
\usepackage{mathrsfs}				
\usepackage{float}
\usepackage{hyperref}

\newcommand{\be}[0]{\begin{equation}}	
\newcommand{\ee}[0]{\end{equation}}

\usepackage{hyperref}
\hypersetup{colorlinks, 
	linkcolor={blue!75!black!80!yellow},
	citecolor={blue!75!black!80!yellow}, 
	urlcolor={blue!75!black!80!yellow}
	}

\makeatletter \renewcommand\@make@capt@title[2]{%
\@ifx@empty\float@link{\@firstofone}{\expandafter\href\expandafter{\float@link}}%
\sffamily{\textbf{#1}}\@caption@fignum@sep#2 }

\usepackage[normalem]{ulem}

\begin{document}


\title{Cavity magnon-polaritons in cuprate parent compounds}


\author{Jonathan B. Curtis}
\email[]{jcurtis@seas.harvard.edu}
\affiliation{John A. Paulson School of Applied Sciences and Engineering, Harvard University, Cambridge Massachusetts 02138 USA}
\affiliation{Department of Physics, Harvard University, Cambridge Massachusetts 02138 USA}
\author{Andrey Grankin}
\affiliation{Joint Quantum Institute, Department of Physics, University
of Maryland, College Park, MD 20742, USA}
\author{Nicholas R. Poniatowski}
\affiliation{Department of Physics, Harvard University, Cambridge Massachusetts 02138 USA}
\author{Victor M. Galitski}
\affiliation{Joint Quantum Institute, Department of Physics, University
of Maryland, College Park, MD 20742, USA}
\author{Prineha Narang}
\affiliation{John A. Paulson School of Applied Sciences and Engineering, Harvard University, Cambridge Massachusetts 02138 USA}
\affiliation{Department of Physics, Harvard University, Cambridge Massachusetts 02138 USA}
\author{Eugene Demler}
\affiliation{Department of Physics, Harvard University, Cambridge Massachusetts 02138 USA}
\affiliation{Institute for Theoretical Physics, ETH Zurich, 8093 Zurich, Switzerland}


\date{\today}

\begin{abstract}
Cavity control of quantum matter may offer new ways to study and manipulate many-body systems.
A particularly appealing idea is to use cavities to enhance superconductivity, especially in unconventional or high-$T_c$ systems.
Motivated by this, we propose a scheme for coupling Terahertz resonators to the antiferromagnetic fluctuations in a cuprate parent compound, which are believed to provide the glue for Cooper pairs in the superconducting phase.
First, we derive the interaction between magnon excitations of the Ne\'el-order and polar phonons associated with the planar oxygens.
This mode also couples to the cavity electric field, and in the presence of spin-orbit interactions mediates a linear coupling between the cavity and magnons, forming hybridized magnon-polaritons.
This hybridization vanishes linearly with photon momentum, implying the need for near-field optical methods, which we analyze within a simple model.
We then derive a higher-order coupling between the cavity and magnons which is only present in bilayer systems, but does not rely on spin-orbit coupling.
This interaction is found to be large, but only couples to the bimagnon operator.
As a result we find a strong, but heavily damped, bimagnon-cavity interaction which produces highly asymmetric cavity line-shapes in the strong-coupling regime. 
To conclude, we outline several interesting extensions of our theory, including applications to carrier-doped cuprates and other strongly-correlated systems with Terahertz-scale magnetic excitations.
\end{abstract}


\maketitle

\section{Introduction\label{sec:intro}}
Using light to control the properties of quantum materials not only holds the potential to realize new and interesting quantum many-body phases~\cite{Choi.2017,Zhang.2017,Heyl.2018,Potirniche.2017,Potter.2017,Oka.2009,Lindner.2011,Kitagawa.2010,Claassen.2017,Bostrom.2020,Else.2016,Bernien.2017,Mivehvar.2017,Chiocchetta.2020,Altman.2002,Bukov.2015}, but may also hold the key to create novel devices and functionalities~\cite{Basov.2017,Cavalleri.2018,Liu.2018,Kennes.2019,Walldorf.2019,Sentef.2017cy,Malz.2019,Gu.2018,Martin.2017,Sternbach.2021,Ebbesen.2016,Orgiu.2015,Gray.2018}.
In most cases, this optical control is achieved by externally applying intense electromagnetic radiation to the system in question~\cite{Mankowsky.2014,Maehrlein.2018,Disa.2020,Budden.2021,Mikhaylovskiy.2020,McLeod.2020,Afanasiev.2021,Qiu.2021,Ron.2020,Buzzi.2021,Kogar.2020,Lovinger.2020,Nova.2019,Cremin.2019,Mitrano.2016,Katsumi.2020,Li.201954b,Shi.20194hj,Tobey.2008,Mikhaylovskiy.2015,Zong.2019,Pashkin.2010,Katsumi.2018,Pilon.2013,Niwa.2019,Matsunaga.2012,Baldini.2020,Sivarajah.2019,Rajasekaran.2016,Dolgirev.2021}.
Recently however, an appealing alternative route has been put forward which bypasses the need for intense external radiation.
Instead, one may attempt to use resonantly-coupled electromagnetic cavities to custom-tailor the properties of the electromagnetic vacuum-fluctuations directly~\cite{Byrnes.2014,Sentef.2018,Curtis.2019,Schlawin.2019,Ashida.2020,Allocca.2019,Raines.2019,Grankin.2020,Kiffner.2019,Basov.2020,Sentef.2020,Kiffner.2019uts,Dehghani.2020,Karzig.2015,Parvini.2020,Zhang.2016nsw,Sentef.2020,Mazza.2019,Scalari.2012,Geiser.2012,Ebbesen.2016,Chiocchetta.2020,Orgiu.2015,Thomas.2019,Raines.2020,Schachenmayer.2015,Hagenmuller.2017,Ashida.2021,Head-Marsden.2020,Rivera.20190wf,Keiser.2021}.

A potentially powerful application of this approach is to use cavities to control antiferromagnetic correlations in a high-$T_c$ cuprate superconductor.
These correlations are believed to underlie many of the exotic, and potentially useful, aspects of the unconventional high-$T_c$ superconductivity in these materials~\cite{Dagotto.1994,Imada.1998,Curty.2002,Peli.2017,Keimer.2015yde}. 
It therefore stands to reason that the ability to optically manipulate, and ultimately enhance, superconductivity in these materials~\cite{Raines.201514,Kennes.2019,Dehghani.2020,Fechner.2016nw9,Mankowsky.2014,Wang.20181st,Okamoto.2016,Cremin.2019} is extremely appealing from both a theoretical and practical perspective, and may pave the way towards realizing room-temperature superconductivity\textemdash a ``holy-grail" of modern condensed matter and material science.

The concept of using cavities to manipulate superconductivity has already been put forward in the context of electron-phonon systems~\cite{Sentef.2018,Hagenmuller.2019,Grankin.2020}.
In this case, it has been proposed that strongly coupling resonant cavities to infrared-active optical phonons~\cite{Woerner.2019,Sumikura.2019,Zhang.2019,Giles.2018} may offer a way to manipulate the pairing between electrons in ``conventional" superconductors.
While indeed this may be a promising avenue~\cite{Thomas.2019} towards achieving cavity-enhanced superconductivity, these systems still remain fundamentally limited by the relatively weak coupling strengths afforded by electron-phonon interaction.
In contrast, while the exact mechanism responsible for superconductivity in cuprates is still unclear, it is widely believed to be driven by interactions of electrons with spin-fluctuations rather than phonons~\cite{Terashige.2019,Emery.1987,Schrieffer.1989,Kane.1989,Dagotto.1994,MacDonald.1988, Chao.1978, Scalapino.1986, Sugawara.1990, Carbotte.1999,Abanov.2003,Lee.2006,Sachdev.2010,Grusdt.2018,Sakai.1997,Mishchenko.2007,Sachdev.2016,Kane.1989,Grusdt.2018,Lee.2003,Lee.2006,Curty.2002,Imada.1998,Emery.1987,Zhang.1988}.
As a result, these systems have a higher ``ceiling" for potential enhancement effects. 
At the same time it is also less clear how cavities may be used to modify these spin-fluctuations. 

Motivated by these considerations, we propose a scheme for realizing strong coupling of a resonant cavity to the spin fluctuations in an antiferromagnetic cuprate parent compound.
Previously, it has been established that optical methods can be used to manipulate magnons in antiferromagnets, either through Raman processes~\cite{Fleury.1968,Chubukov.1995,Zhao.2004,Devereaux.1995,Devereaux.1999,Wang.2018,Lee.1996}, or ultrafast optical methods~\cite{Wang.2018,Disa.2020,Seifert.2019,Bolens.2018,Maehrlein.2018,Fechner.2016nw9,Grishunin.2021,Sivarajah.2019,Kampfrath.2011,Yamaguchi.2010,Reid.2015,Nishitani.2010}.
However, most of these protocols are difficult to adapt to the cavity setting due to the fact that they essentially rely on nonlinear or strongly off-resonant processes~\cite{Parvini.2020}.
Unlike their ferromagnetic counterparts, which reside at GHz frequencies~\cite{Xu.2020,Zhang.2014vxb,Li.2019,Neuman.2020}, magnons in an antiferromagnetic typically reside in the 1-10 THz regime, owing to their larger superexchange energy scales~\cite{Keimer.1993}.
This presents additional challenges due to the notorious difficulties in Terahertz engineering which are only recently being overcome~\cite{Yen.2004,Benea-Chelmus.2020,Zhang.2018jbq,Woerner.2019,Zhang.2016nsw,Klingler.2016,Mandal.2020,Jiang.2016,Zhang.2019,Wu.2020,Schalch.2019,Sloan.2019,Maissen.2014ta,Scalari.2012,Geiser.2012,Ebbesen.2016,Orgiu.2015,Thomas.2019,MacNeill.2019,Bayer.2017,Paravicini-Bagliani.2019}.
In this Article we theoretically address these challenges, thereby taking the first crucial step towards achieving cavity control of high-temperature superconductivity in cuprates.

\begin{figure}
    \centering
    \includegraphics[width=\linewidth]{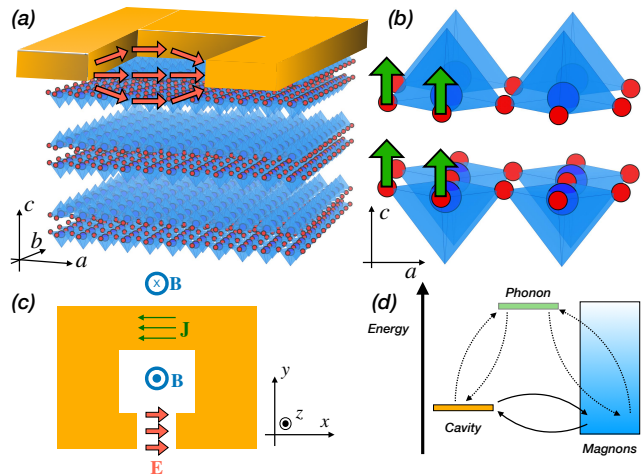}
    \caption{Overview of our proposal for forming cavity magnon-polaritons in insulating cuprates. 
    (a) Depiction of the combined cavity-sample system. 
    We consider placing the resonator on top of the cuprate (YBCO) sample, which is quasi-two dimensional and consists of layers of copper-oxide bilayers (for clarity we only depict the planar copper and oxygen atoms). 
    The resonator produces a strong electric field (red arrows) which penetrates into the sample and couples to the magnons via the scheme outlined in Sec.~\ref{sec:model}.
    (b) The phonon mode we consider in this paper, shown in a simplified depiction of the YBCO unit cell, is associated to the infrared-active motion of planar oxygens along the $c$-axis and strongly couples to both the cavity photons and the magnons in the cuprate.
    (c) Depiction of a typical split-ring resonator, which we use to model the Terahertz cavity in this work.
    Current flows through the metallic ring, inducing a magnetic field, as well as producing an electric field in the capacitive gap. 
    (d) Cavity photon mode (gold) is constructed to lie close to the bottom of the magnon band (blue continuum), which are the two degrees we wish to couple to each other. 
    The magnon-cavity coupling is mediated by virtual phonons, which both systems couple to directly. 
    Because the phonon mode in question is higher in frequency than the relevant magnons and cavity, it can be adiabatically eliminated, yielding the desired cavity-magnon coupling. }
    \label{fig:fig1}
\end{figure}

This leads us to our central proposal of using Terahertz resonators to strongly couple to the magnons in a cuprate system, which is schematically depicted in Fig.~\ref{fig:fig1}.
Specifically, we propose two suitable microscopic mechanisms by which magnons in the insulating cuprate may be coupled to cavity photons.
Both of these employ an infrared-active phonon mode~\cite{Sivarajah.2019}, depicted in Fig.~\ref{fig:fig1}(b), as an intermediary between the electric field of the cavity and the electron spins, shown schematically in Fig.~\ref{fig:fig1}(d).
In the first scenario, we examine how in the presence of spin-orbit coupling~\cite{Coffey.1991,Bonesteel.1993,Shekhtman.1992,Koshibae.1993,Shekhtman.1993,Koshibae.1993vbj,Wu.2005,Atkinson.2020,Raines.2019} this phonon mode can lead to a linear coupling between the cavity photons and spin-waves~\cite{Cao.2015,Juraschek.2017,Talbayev.2011,Katsura.2007,Shuvaev.2010,Falko.2006,Lee.2020f7,Chaudhary.20200jr,Malz.2019}.
We find that while this does lead to a linear coupling, in the absence of inversion-symmetry breaking~\cite{Zhao.2017} this coupling vanishes in the far-field limit and the magnons decouple from the electric field.
While this poses a problem for conventional Terahertz optics, since Terahertz resonators are localized well within the sub-wavelength regime, this will not fundamentally impede the coupling to a resonant cavity. 
Inspired by recent advances in Terahertz near-field optics~\cite{Sun.20205op,Zhang.2018jbq,Bitzer.2009,Lu.2020,Jiang.2016,Zhang-thesis.2019,Basov.2016,Schalch.2019,Stinson.2014,Fei.2011,Sunku.2018,Maissen.2014ta}, we explore a particular model in which a strong near-field coupling is realized.

In the second scenario, we examine how the same infrared-active phonon mode couples to the scalar spin-exchange even in the absence of spin-orbit interactions. 
In a bilayer system, this mode couples linearly to the spin exchange interaction once one takes into account the buckled nature of the equilibrium structure~\cite{Normand.1996,Sakai.1997}.
However, while this mode couples linearly to the exchange, it still only couples at quadratic order in the spin-waves.
While this coupling doesn't lead to the formation of polaritons, we show that nevertheless this does lead to a strong coupling between the photons and cavity, and may be promising in the future since it more naturally allows for controlling the correlations which are moderated by the spin-waves~\cite{Erlandsen.2019,Erlandsen.2020,Grankin.2020,Kennes.2017,Babadi.2017,Knap.2016,Murakami.2017}. 
Both of these processes are argued to be within the reach of current experiments. 
A summary of the main results, and which section they are discussed in, is encapsulated in Table~\ref{tab:summary}. 

\begin{table}
\begin{center}
\begin{tabular}{ |c|c|c|c| }
 \hline
 \textbf{Section} & \ref{sec:mono-soc} & \ref{sec:bi-soc} & \ref{sec:bimagnon}  \\ 
 \hline 
 \textbf{Structure} & monolayer & bilayer & bilayer  \\
 \textbf{Mechanism} & spin-orbit & spin-orbit  & buckling \\
 \textbf{Order} & linear & linear  & quadratic \\
 \textbf{Cavity type} & near-field & near-field & conventional \\
 \hline
\end{tabular}
\end{center}
\caption{Summary of the coupling schemes considered in this work and the section where they are discussed. 
For each, we have indicated the structure of the system (monolayer vs. bilayer) considered, the physical mechanism underlying the coupling (spin-orbit interaction vs. buckling modulation), the order of magnon-operator coupled to (linear coupling to single magnon operator or quadratic coupling to bi-magnon operator), and whether a near-field or conventional cavity is needed for experimental detection.}
\label{tab:summary}
\end{table}

The remainder of this paper is structured as follows: in Sec.~\ref{sec:model} we review the equilibrium structure of the model cuprate system and its interactions with lattice vibrations and spin-orbit effects. 
Next, in Sec.~\ref{sec:polaritons} we focus our attention on a resulting linear magneto-electric coupling, and show how it can lead to hybridized cavity magnon-polaritons.
After examining the linear-coupling, we proceed to Sec.~\ref{sec:bimagnon}, where we examine a quadratic coupling between the cavity photons and bimagnons also present in our model, and argue it can also give rise to strong-coupling signatures between the cavity-photons and bimagnons. 
Finally, we conclude with a discussion in Sec.~\ref{sec:conclusion}, wherein we outline future interesting directions and applications of our work.

\section{Model\label{sec:model}}

We begin with the nature of spin orbit coupling in the cuprates. 
We will mainly focus on modeling bilayer cuprate systems, as realized by the YBa$_2$Cu$_2$O$_6$ (YBCO) family of compounds.
However, it is often qualitatively similar, but technically simpler, to consider a model system which consists of a single monolayer with easy-plane anisotropy.
The details of the derivation of the easy-plane toy model will be relegated to App.~\ref{app:EP}, though it will also straightforwardly follow as a limiting case of decoupled bilayers.

\subsection{Spin-Orbit Coupling\label{sub:soc}} 
We will first focus on the case of a bilayer system, such as YBCO.
Importantly, in the compound YBCO there are two copper oxide planes in each unit cell which are related to each other by horizontal mirror plane symmetry.
Therefore, overall inversion symmetry is preserved, but can still act non-trivially on each layer~\footnote{This is aside from the possible reports of inversion symmetry breaking~\cite{Zhao.2017,Viskadourakis.2015,Mukherjee.2012}, which we do not address here.}. 

To lowest order, we expect that at half-filling the low-energy Hamiltonian for the system should roughly map onto a nearest-neighbor Heisenberg model with weak inter-layer coupling~\cite{Bonesteel.1993} of the form 
\begin{equation}
    \label{eqn:ybco-bare-hamiltonian}
    \hat{H}_0 = \sum_{\ell} \sum_{<j,k>} J_0 \hat{\bm{S}}_{j\ell}\cdot\hat{\bm{S}}_{k\ell} + \sum_{j} J_{ud} \hat{\bm{S}}_{ju} \cdot \hat{\bm{S}}_{jd}
\end{equation}
where the vector operator $\hat{\bm{S}}_{j\ell}$ describes the spin-$\frac12$ moment located at site $\mathbf{R}_j$ of layer $\ell = u,d$, and $<j,k>$ indicates that $j$ and $k$ are nearest-neighbor sites within the $ab$-plane. 
The inter-layer coupling $J_{ud}$ involves the hopping along the $c$-axis and is typically two orders of magnitude than the intra-layer processes~\cite{Bonesteel.1993}.

So far, this model neglects spin-orbit effects which, though small, are important at the Terahertz scale, as they set the scale for the magnon gap~\cite{Keimer.1993,Bonesteel.1993}.
To obtain the spin-orbit coupling corrections to super-exchange~\cite{Anderson.1959,Dzyaloshinsky.1958,Moriya.1960}, we start with a single-band bilayer Hubbard model with spin-orbit coupled hopping $i\bm\lambda\cdot\bm\sigma$~\cite{Das.2013}.
The relevant electronic Hamiltonian is
\begin{equation}
    \label{eqn:rashba-hubbard}
    \hat{H}_{\rm el} = - \sum_{j\bm{\delta} \ell} \sum_{\alpha\beta}\hat{c}_{j+\bm{\delta},\ell \alpha}^\dagger \left[ t + i \bm{\lambda}_{\bm{\delta}\ell} \cdot \bm{\sigma} \right]_{\alpha\beta} \hat{c}_{j \ell \beta} + U \sum_{j\ell} \hat{n}_{j\ell \uparrow}\hat{n}_{\ell \downarrow}.
\end{equation}
Here $\hat{c}_{j\ell\sigma}$ annihilates an electron with spin $\sigma$ from lattice site $\mathbf{R}_j$ in layer $\ell$.
Throughout this work, repeated spin indices are implicitly summed over.
We have also introduced the nearest-neighbor lattice vectors $\bm{\delta}$, and used the shorthand $j+\bm{\delta}$ to indicate the lattice site located at real-space coordinate $\mathbf{R}_j + \bm{\delta}$.
We assume isotropic spin-symmetric hopping, denoted by the hopping integral $t$, while $\bm{\lambda}_{\bm\delta \ell}$ is the (smaller) spin-orbit correction to the hopping, which is different in each bond direction $\bm{\delta}$ and also depends on the layer index $\ell$.
As a rough estimate, we will assume typical values of $t \sim \SI{400}{\meV}$ and $U \sim\SI{4.2}{\eV}$~\cite{Dagotto.1994}.

By time-reversal symmetry, we have that $t$ and $\bm{\lambda}$ are real, and the constraint of Hermiticity requires that $\bm{\lambda}_{\bm{\delta}} = -\bm{\lambda}_{-\bm{\delta}}$.
For the case of a monolayer, inversion symmetry then requires $\bm{\lambda}_{\bm \delta} = \bm{\lambda}_{-\bm{\delta}}$, and thus the spin-orbit coupling must vanish.
However, the crucial observation is that in the case of YBCO, inversion symmetry only constrains the spin-orbit coupled hopping to obey $\bm{\lambda}_{-\bm{\delta} u} =  \bm{\lambda}_{\bm {\delta}d}$, which allows for non-zero spin-orbit coupling on each layer.
It is worth explicitly reminding that the spin-orbit coupling texture $\bm \lambda$ is formally a pseudo-vector.
A natural choice for the form of the spin-orbit coupling is the Rashba-like pattern 
\begin{equation}
    \bm{\lambda}_{\bm{\delta}\ell} = (-1)^{\ell} \lambda {\bf e}_z\times\bm{\delta},
\end{equation}
which assumes the value $\lambda {\bf e}_y$ on the upper-layer $x$-bonds and $-\lambda {\bf e}_x$ on the upper-layer $y$-bonds. 
Following Ref.~\onlinecite{Atkinson.2020}, we take as a rough estimate the spin orbit interaction $\lambda \sim \SI{10}{\meV}$.

Microscopically, this spin-orbit coupling naturally emerges if one considers the $dp\sigma$ bonds between the planar oxygen atoms and the excited $t_{2g}$ crystal-field states of the copper ions (in particular, the $d_{xz}$ and $d_{yz}$ bands)~\cite{Coffey.1991,Bonesteel.1993,Shekhtman.1992,Koshibae.1993,Shekhtman.1993,Koshibae.1993vbj,Wu.2005,Atkinson.2020,Raines.2019,Normand.1996}.
In a bilayer system like YBCO, local crystal field gradients on each layer cause a uniform buckling of the planar oxygens in the copper oxide planes towards the interior of the bilayers~\cite{Bonesteel.1993,Coffey.1991,Shekhtman.1992,Atkinson.2020,Normand.1996,Raines.2019,Das.2013}.
This displaces the oxygens by about $\SI{.22}{\angstrom}$ from the copper oxide plane, leading to a non-linear bond angle.
The Rashba spin-orbit coupling correction then emerges at linear order from the atomic $\mathbf{L}\cdot\mathbf{S}$ coupling, which linearly couples the $e_g$ ground-state to the $t_{2g}$ excited states.
Recently, effects of similar spin-obit couplings have been seen in the dynamics of charge carriers in hole-doped cuprates YBCO~\cite{Harrison.2015,Atkinson.2020} and Bi-2212~\cite{Gotlieb.2018,Raines.2019}, reinforcing this theoretical argument.

It is now straightforward to obtain the spin-orbit corrections to Hamiltonian~\eqref{eqn:ybco-bare-hamiltonian} due to spin-orbit coupling in equation~\eqref{eqn:rashba-hubbard}.
As we derive in Appendix~\ref{app:low-energy-hubbard}, one arrives at the low-energy ``Rashba bilayer" spin-model 
\begin{multline}
        \label{eqn:spin-model}
\hat{H}_{\rm BL} = \sum_{j\in A\ell \bm{\delta}} J_0\hat{S}_{j\ell}\cdot \hat{S}_{j+\bm \delta \ell}  + \Gamma^{ab}_{\bm {\delta}}  \hat{S}^a_{j\ell} \hat{S}^b_{j+\bm \delta \ell} +  \mathbf{D}_{\bm \delta \ell} \cdot \hat{\mathbf{S}}_{j\ell}\times\hat{\mathbf{S}}_{j+\bm{\delta}\ell} \\
+ \sum_{j} J_{ud}\hat{\mathbf{S}}_{ju}\cdot\mathbf{\hat{S}}_{jd} ,
\end{multline}
with intra-layer exchange tensors 
\begin{subequations}
\label{eqn:rashba-exchange-constants}
\begin{align}
&J_0 = 4\frac{t^2 -\lambda^2}{U} \\
& \Gamma_{\bm \delta}^{ab} \equiv \Gamma \left(\mathbf{e}_{z}\times \bm{\delta} \right)_a  \left(\mathbf{e}_{z}\times \bm{\delta} \right)_b =  \frac{8\lambda^2}{U} \left(\mathbf{e}_{z}\times \bm{\delta} \right)_a  \left(\mathbf{e}_{z}\times \bm{\delta} \right)_b \\
& \mathbf{D}_{\bm \delta\ell} \equiv D_\ell \mathbf{e}_z \times \bm \delta = -(-1)^{\ell}\frac{8t\lambda}{U} \mathbf{e}_{z}\times \bm{\delta}.
\end{align}
\end{subequations}
In order to avoid double-counting, we have implemented the sum over bonds as a sum over sites in the A sublattice, and a sum over its neighbors (which reside in the B sublattice).

We are now interested in coupling this spin system to the optical field of an electromagnetic cavity.
We will outline a few possible mechanisms to achieve this coupling, all of which exploit electron-phonon interactions.
Within the context of ultrafast optics, phono-magnetic interactions are by now a well established way of coupling Terahertz optical fields to electronic spins~\cite{Afanasiev.2021,Potter.2013,Son.2019,Padmanabhan.2020,Juraschek.2017,Juraschek.2020,Mikhaylovskiy.2020,Mikhaylovskiy.2015,Maehrlein.2018,Juraschek.2020i8}.
The combination of a relatively large electric dipole matrix element and strong electron-phonon interaction~\cite{Stephan.1997} at the appropriate energy scale also makes this a promising route for engineering linear coupling of spins to the cavity electromagnetic field.
However, before proceeding, it is worth briefly commenting that other mechanisms, such as direct magnetic-dipole coupling~\cite{Mukai.2014,Yuan.2020,Klingler.2016,Grishunin.2021,Sivarajah.2019,Kampfrath.2011,Yamaguchi.2010,Reid.2015,Nishitani.2010,Mandal.2020,Yuan.2017,Wu.2020,Sloan.2019,Camley.1980,Liu.2020917,MacNeill.2019,Neuman.2020}, or more complicated electric-dipole couplings~\cite{Tanabe.1965,Moriya.1968,Fleury.1968,Shastry.1990,Katsura.2004,Jia.2006,Jia.2007,Bolens.2018,Seifert.2019,Parvini.2020,Bulaevskii.2008,Juraschek.2021tyfp} may also be potentially fruitful but lie outside the scope of this paper and will be discussed in future works.
We now explore one particular phonon mode and show how it can be used to obtain a coupling between the spin system and the cavity mode.

\subsection{Polar Phonon Coupling\label{sub:phonon}}
Typically, a polar phonon mode cannot couple linearly to the spin degrees of freedom in a centrosymmetric system due to inversion symmetry.
We will investigate two ways around this obstruction.
The first is through spin-orbit interactions~\cite{Katsura.2007,Cao.2015,Juraschek.2017,Talbayev.2011,Shuvaev.2010,Falko.2006,Lee.2020f7,Chaudhary.20200jr}.
This mechanism is generic to both the monolayer and bilayer systems, and generates both linear and quadratic couplings.
The second mechanism is specific to the bilayer system and only generates quadratic couplings, but is expected to be much stronger than the spin-orbit interaction~\cite{Opel.1999}. 
This interaction can be understood as arising from linearizing the standard bi-quadratic $\sim Q^2 \mathbf{S}\cdot\mathbf{S}$ coupling of the polar mode around the buckled equilibrium structure, which is allowed only in the bilayer system. 

For the purpose of specificity, we will focus on the well-known infrared-active $B_{1u}$ mode which is located at a frequency of roughly $\SI{9}{\THz}$ ($\SI{41}{\meV}, 320 \si{cm}^{-1}$)~\cite{Liu.2020wx,Kress.1988,Liu.1988,Munzar.1999,Mankowsky.2014,Homes.1995,Devereaux.1995,Devereaux.1999,Devereaux.2004,Johnston.2010,Humlicek.1993,Henn.1997,Pashkin.2010,Cooper.1993,Normand.1996,Pashkin.2010,Fechner.2016nw9,Kovaleva.2004} in YBCO.
This mode corresponds to a uniform, in phase $c$-axis vibration of all of the planar oxygen atoms in both bilayers.
The bare phonon dynamics is modeled by the Hamiltonian 
\begin{equation}
\label{eqn:phonon}
\hat{H}_{\rm ph} = \sum_{j\in A,\bm \delta} \frac{1}{2M_{\rm ph}}\hat{P}_{j,j+\bm \delta}^2 + \frac12 M_{\rm ph} \Omega_{\rm ph}^2 \hat{Q}_{j,j+\bm \delta}^2 - Z_{\rm ph} e \hat{E}^z_{j,j+\bm \delta}\hat{Q}_{j,j+\bm \delta},
\end{equation}
where to avoid double counting we have instituted the sum over all bonds by introducing the two sublattices A and B and summing over each site in A and all of its neighboring sites at $j+\bm \delta$, which are in sublattice B.
Here $\hat{Q}_{j,j+\bm \delta}$ describes the phonon displacement of the oxygen residing on the bond connecting sites $j\in A$ and $j+\bm \delta \in B$, and $\hat{P}_{j,j+\bm \delta}$ is the corresponding canonically conjugate momentum. 
In addition to the resonance frequency of $\Omega_{\rm ph} \sim \SI{41}{\meV}$, we also have introduced the phonon effective mass $M_{\rm ph} \sim 8 M_{\rm O}$, and Born effective charge $Z_{\rm ph} \sim 8 Z_{\rm O}$, with $Z_{\rm O}, M_{\rm O}$ the effective charge and mass of one of the planar oxygens.  

To capture the interaction with the electrons, we also introduce the electron-phonon coupling Hamiltonian 
\begin{multline}
\label{eqn:el-ph-int}
\hat{H}_{\rm el-ph} = \sum_{j\in A,\bm \delta, \ell} c^\dagger_{j+\bm{\delta}\ell}\left(i\kappa \hat{Q}_{j,j+\bm{\delta}} \mathbf{e}_z\times \bm{\delta}\cdot \bm\sigma \right)c_{j\ell} + \textrm{h.c.} \\ 
+ \sum_{j\in A,\bm \delta, \ell}  - \alpha (-1)^\ell \hat{Q}_{j,j+\bm{\delta}} c^\dagger_{j+\bm{\delta}\ell}c_{j\ell} + \textrm{h.c.}, 
\end{multline}
where the first term describes the change in the spin-orbit hopping due to the phonon, while the second term describes the change in transfer intensity, and is only present in the buckled bilayer structure. 

We now perform a rough estimation of the sizes of the electron-phonon coupling constants, based on the known equilibrium properties. 
To estimate the spin-orbit parameter $\kappa$, we observe that in equilibrium the spin-orbit exchange in the bilayer system is roughly $\lambda\sim \SI{10}{\meV}$~\cite{Atkinson.2020}.
For small displacements, this should be roughly proportional to the equilibrium oxygen displacement $Q_{\rm O} \sim \SI{.22}{\angstrom}$.
We therefore extrapolate that the constant $\kappa \sim \lambda /Q_{\rm O} \sim \SI{45}{\meV\per\angstrom}$.

The parameter $\alpha$ is more difficult to estimate accurately. 
We will use the rough estimates provided in Ref.~\onlinecite{Normand.1996,Sakai.1997}, where it is estimated that $2\alpha/t  \sim 5/\SI{.22}{\angstrom}$ and therefore this coupling can be quite large, with effective spin-phonon coupling constant $8\alpha t /U  \sim \SI{2.7}{\eV\per\angstrom}$.

We now treat the combined Hubbard-phonon-photon system in the half-filling manifold by following the usual procedure. 
We will only consider linear corrections in the phonon amplitude, and therefore we can treat the operator as a c-number for the purposes of the Schrieffer-Wolf transformation~\cite{Stephan.1997}.
Furthermore, we will only select the largest interactions to consider in more depth, though all of the expected terms can be found in Appendix~\ref{app:low-energy-hubbard}.
To linear order in the displacement, we find that Eq.~\eqref{eqn:spin-model} is modified to include the spin-phonon coupling  
\begin{multline}
\label{eqn:spin-phonon}
H_{\rm sp-ph} = \sum_{j\ell \bm{\delta}} \frac{8t \alpha}{U}  (-1)^{\ell } \hat{Q}_{j,j+\bm\delta}  \hat{\mathbf{S}}_{j\ell}\cdot \hat{\mathbf{S}}_{j+\bm{\delta}\ell} \\
+\frac{8t\kappa}{U} \hat{Q}_{j,j+\bm\delta} \left(\mathbf{e}_z\times \bm \delta\right)\cdot \hat{\mathbf{S}}_{j\ell}\times \hat{\mathbf{S}}_{j+\bm{\delta}\ell}.  
\end{multline}
We note that if we were to ignore the buckling due to the bilayer structure, or simply considered a monolayer material, inversion symmetry would force the first term to vanish, but still would allow for the second term.
We roughly estimate that $4t/U \sim .38, \kappa \sim \SI{45}{\meV \per \angstrom}$, so that the spin-phonon coupling constant is approximately $8t\kappa/ U \sim \SI{34}{\meV\per\angstrom}$. 
As roughly quoted above, the larger constant is $8\alpha t /U  \sim \SI{2.7}{\eV\per\angstrom}$.

If we approximate the phonon frequency to be large compared to the frequency of the spin-fluctuations of interest, we can treat the phonon in the Born-Oppenheimer approximation.
This is acceptable for magnetic zone-center spin-waves, which have frequencies of order $2-3\ \si{\meV}$, however in general the magnon bandwidth is still much larger than the phonon resonance frequency and in this case, a more involved treatment of the full spin-phonon-photon model is needed.
Since the interaction with the cavity is dominated by the zone-center magnons (see discussion in Sec.~\ref{sub:near-field}), we proceed to make the Born-Oppenheimer approximation obtain a direct magnetoelectric coupling 
\begin{multline}
    \label{eqn:me-slow-phonon}
    H_{\rm int} = \sum_{j\ell \bm{\delta}} g'(-1)^\ell  \hat{E}^z_{j,j+\bm\delta}  \hat{\mathbf{S}}_{j\ell}\cdot \hat{\mathbf{S}}_{j+\bm{\delta}\ell} \\
+g \hat{E}^z_{j,j+\bm\delta} \left(\mathbf{e}_z\times \bm \delta\right)\cdot \hat{\mathbf{S}}_{j\ell}\times \hat{\mathbf{S}}_{j+\bm{\delta}\ell},  
\end{multline}
with the magnetoelectric coupling constants $g,g'$ given by   
\begin{subequations}
\begin{align}
& g= \frac{Z_{\rm ph}e }{M_{\rm ph}\Omega^2_{\rm ph}}\frac{8t\kappa}{U} \\
& g'= \frac{Z_{\rm ph}e }{M_{\rm ph}\Omega^2_{\rm ph}}\frac{8t\alpha}{U}.
\end{align}
\end{subequations}
Using a rough crystal-field approximation, we estimate that $Z_{\rm ph} = 2e$, $M_{\rm ph}= 8 \si{\amu}$, and $\Omega_{\rm ph} = \SI{10}{\THz}$.
Therefore, the static polarizeability of the phonon mode in question is $Z_{\rm ph} e / (M_{\rm ph} \Omega_{\rm ph}^2 ) \sim 1.3\times 10^{-4} \si{\angstrom}/\si{\kV\per\cm}$.
This corresponds to a value of $g \sim  4.4 \times 10^{-3} \si{\meV}/\si{\kV\per\cm}$.
An electric field strength on the order of $E^z\sim \SI{100}{\kilo\volt\per\cm}$ would then lead to a change in the strength of the Dzyaloshinskii-Moriya (DM) interaction by an amount of order $0.4\, \si{\meV}$ or roughly 5\% of the equilibrium value.
We also find that $g' = (\alpha/\kappa) g \sim 80 g$ can be quite large, though there is a notable degree of uncertainty in the size of our estimate of the coupling constant $\alpha$.

Altogether, we are now tasked with studying the dynamics of the full effective Hamiltonian 
\begin{multline}
    \label{eqn:full-me-ham}
    H = \sum_{j\in A \ell \bm{\delta}} J^{ab}_{\bm {\delta}}  \hat{S}^a_{j\ell} \hat{S}^b_{j+\bm \delta \ell} + \mathbf{D}_{\bm \delta \ell}\cdot \hat{\mathbf{S}}_{j\ell}\times\hat{\mathbf{S}}_{j+\bm{\delta}\ell}   \\
    + \sum_{j\in A,B}J_{ud} \hat{\mathbf{S}}_{ju} \cdot\hat{\mathbf{S}}_{jd}\\
  + \sum_{j\in A \ell \bm{\delta}} g \hat{E}^z_{j,j+\bm \delta} (\mathbf{e}_z \times \bm\delta)\cdot \hat{\mathbf{S}}_{j\ell}\times\hat{\mathbf{S}}_{j+\bm{\delta}\ell} +  g' (-1)^\ell \hat{E}^z_{j,j+\bm \delta} \hat{\mathbf{S}}_{j\ell}\cdot\hat{\mathbf{S}}_{j+\bm{\delta}\ell}. 
\end{multline}
Here we have included the weak inter-layer superexchange $J_{ud}$, which is known to play a crucial role in stabilizing the observed N\'{e}el order~\cite{Bonesteel.1993}.
Furthermore, in the absence of spin-orbit considerations, it is important to include an interlayer coupling term, otherwise for a uniform $E^z$ the dynamics of the $g'$ interaction will actually be trivial since it will commute with the unperturbed dynamics.
We have also implemented the convention that the sum over intralayer bonds is carried out by introducing the $A$ and $B$ sublattices and summing over sites only in the $A$ sublattice, along with each nearest-neighbor bond of that site (which is a site lying in the $B$ sublattice). 
Before examining the consequences of the magnetoelectric coupling, we will review the equilibrium spin-wave dynamics of Hamiltonian~\eqref{eqn:full-me-ham}.
In addition, since the full spin dynamics of Hamiltonian~\eqref{eqn:full-me-ham} is rather complicated, we will also analyze a stand-in easy-plane model which, when possible, we will use to simplify the analysis.
This Hamiltonian can be found below in Eq.~\eqref{eqn:ep-ham} and essentially captures the most important aspect of the full spin-orbit interaction above, which is that it favors in-plane Neel order.

\section{Equilibrium Spin-Wave Spectrum\label{sec:eq-spin-waves}}

We now quickly review the equilibrium spin-wave dispersion relations.
We will first study the dynamics of the simplified monolayer easy-plane antiferromagnet before proceeding on to study the full bilayer Rashba model~\eqref{eqn:full-me-ham}, which will end up bearing a number of similarities to the monolayer toy model. 

\subsection{Easy-Plane Toy Model\label{sub:ep-eq}}

In order to gain intuition, we begin by studying a simplified model of a single copper oxide plane with Hamiltonian 
\begin{equation}
\label{eqn:ep-ham}
    H_{\rm EP} = \sum_{j\in A, \bm \delta } J_0 \hat{\mathbf{S}}_j \cdot \hat{\mathbf{S}}_{j+\bm \delta} - \Gamma \hat{S}^z_j \hat{S}^z_{j+\bm \delta},
\end{equation}
which has a uniaxial anisotropy $\Gamma> 0$.
We emphasize again, this model is meant to serve as a conceptual aid in understanding the behavior of the more complicated mode given above in Eq.~\eqref{eqn:full-me-ham}.
The mean-field ground state of this system has N\'eel order which lies in the $a-b$ plane; we take it to lie at an angle $\theta$ with respect to the crystalline $a$ axis.
We now pass to a right-handed coordinate system, with unit vectors $\mathbf{e}_1,\mathbf{e}_2,\mathbf{e}_3$, such that $\mathbf{e}_3$ is aligned with the Neel order and $\mathbf{e}_1 = \mathbf{e}_z$.
For details, we refer the interested reader to Appendix~\ref{app:EP}. 

We now perform the standard Holstein-Primakoff expansion for the spin operators in this coordinate system.
These are compactly collected into a four-component bosonic Nambu spinor $\Psi_{\bf p} = (a_{\bf p},b_{\bf p}, a_{-\bf p}^\dagger, b_{-\bf p }^\dagger )^T$.
We also introduce the Nambu-space Pauli-matrices $\bm \tau$ and sublattice-space Pauli matrices $\bm \zeta$.
Expanding the spin-wave Hamiltonian to quadratic order we then find 
\begin{multline}
    H_{\rm EP}  = \frac12 \sum_{\bf p} \Psi_{\bf p}^\dagger \left[ 2J_0 - \Gamma \gamma_{\bf p}^s \zeta_1 + (2J_0 - \Gamma) \gamma_{\bf p}^s \zeta_1 \tau_1 \right]\Psi_{\bf p},
\end{multline}
where $\gamma_{\bf p}^s$ is the cubic harmonic of (extended) $s$-wave symmetry 
\begin{equation}
    \gamma_{\bf p}^s = \frac12 \left( \cos p_x + \cos p_y \right).
\end{equation}
This can be diagonalized by standard unitary and Bogoliubov transformations (see Appendix~\ref{app:EP}).
We find two dispersion relations indexed by their sublattice quantum number $\zeta = \pm 1$ 
\begin{equation}
\label{eqn:EP-dispersion}
    \Omega_{\bf p,\zeta} = \sqrt{ (2J_0 - \zeta \Gamma\gamma_{\bf p}^s )^2  - (2J_0 - \Gamma)^2 (\gamma_{\bf p}^s)^2 }.
\end{equation}
The $\zeta = -1$ mode is gapped to a frequency $\sqrt{8J_0\Gamma}$, while the $\zeta= +1$ mode remains a gapless Goldstone mode, with spin-wave velocity $v\sim \sqrt{2J(2J-\Gamma)} $.
In the absence of interlayer exchange or further in-plane anisotropies, true long-range N\'eel order is then destroyed at finite temperatures in accordance with the Mermin-Wagner theorem.

\begin{figure}
    \centering
    \includegraphics[width=\linewidth]{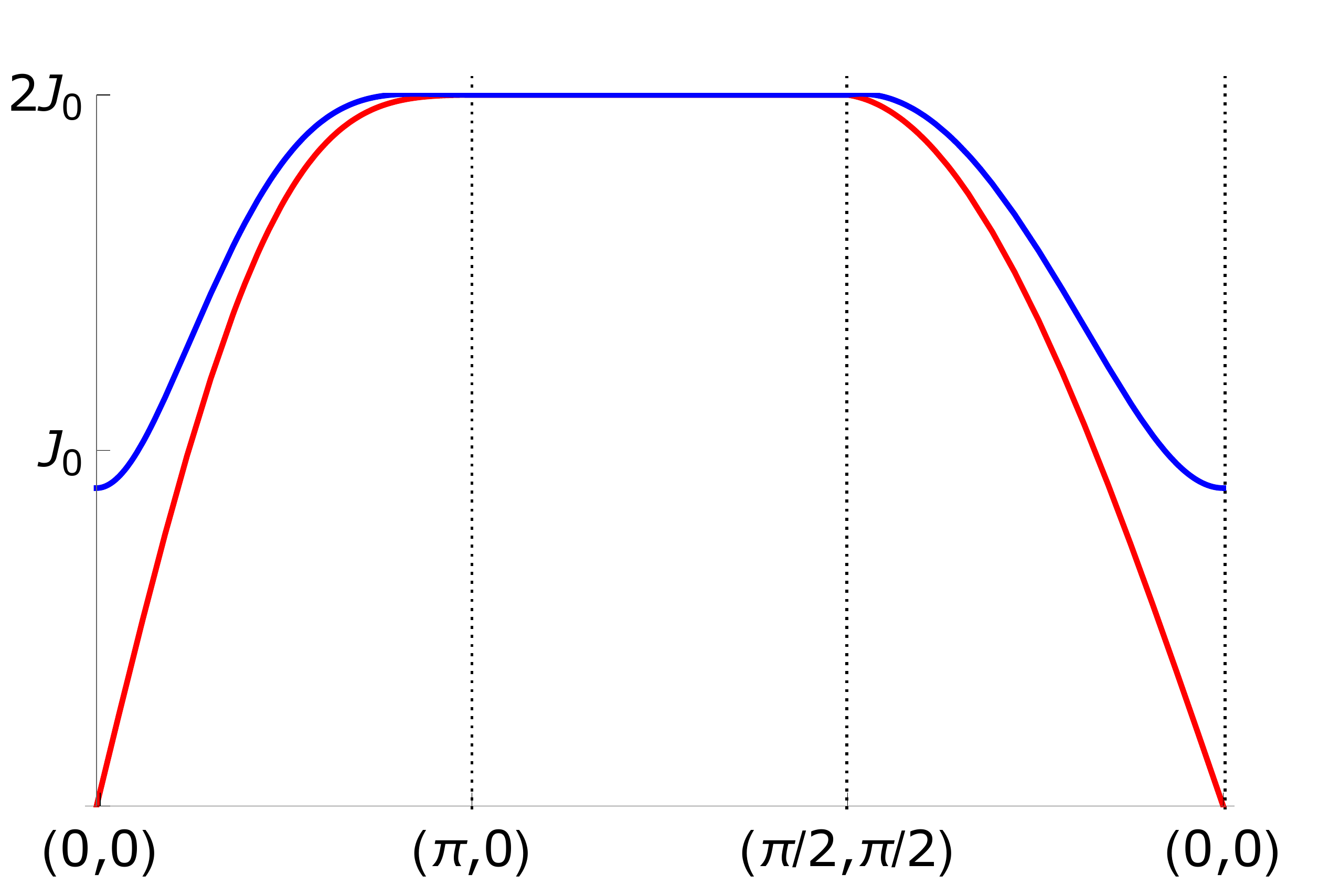}
    \caption{Dispersion of two spin-wave bands along the contour in the reduced magnetic Brillouin zone shown in the inset. 
    This is shown for an exaggerated anisotropy of $\Gamma \sim .1 J_0$, which results in a large spin-wave gap for one of the modes (blue).
    In reality, the anisotropy is expected to be much smaller, since the gap scales with the square root of the anisotropy.}
    \label{fig:EP-eq-dispersion}
\end{figure}

\subsection{Full Bilayer Rashba Model\label{sub:rashba-eq}}
We now proceed to consider the full bilayer Hamiltonian of Eq.~\eqref{eqn:spin-model}.
Following Ref.~\cite{Bonesteel.1993}, we make the ansatz that the ground state is a N\'eel order which lies in the $a-b$ plane (again at angle $\theta$ with respect to the $a$ axis) and that is opposite on the two layers.
The motivation for the easy-plane N\'eel order is that all anisotropies in the Hamiltonian~\eqref{eqn:full-me-ham} lie within the $a-b$ plane.
Specifically, the Rashba spin-orbit coupling results in an {\bf easy-axis} anisotropy which favors N\'eel ordering along the $b$-axis for $x$-oriented bonds, and $a$-axis ordering along the $y$-oriented bonds.
While this is clearly frustrated between ordering along $a$ or $b$, it is clear that no anisotropies favor $c$-axis order, and therefore the N\'eel order in the ground state lies in the $a-b$ plane~\footnote{Due to frustration at the mean-field level, the ground-state remains invariant under in-plane rotations of the N\'eel vector, even though spin-rotation symmetry is explicitly broken by spin-orbit interactions. 
Therefore, whereas linear spin-wave theory predicts that one mode will remain gapless as a Goldstone mode, this is in fact a pseudo-Goldstone mode which must ultimately acquire a gap through an order-by-disorder mechanism.}.

Since the N\'eel order is assumed to lie in the $a-b$ plane, we adopt the same axes as in the easy-plane calculation, and again expand to quadratic order in the Holstein-Primakoff bosons.
For details of the expansion, see Appendix~\ref{app:bilayer}.
Unlike the previous easy-plane case, there are now two layers and therefore the spin-wave operators now carry a bilayer quantum-number.
This brings the bosonic Nambu spinor $\Psi_{\bf p}$ up to eight-components, and introduces a new set of Pauli matrices, $\bm l$, which act on the layer space. 
The full spin-wave Hamiltonian is $H = \frac12 \sum_{\bf p} \Psi_{\bf p}^\dagger \mathbb{M}_{\mathbf{p}} \Psi_{\bf p}$, with $8\times 8$ bilinear matrix 
\begin{widetext}
\begin{equation}
\label{eqn:eq-rashba-spin-wave}
\hat{\mathbb{M}}_{\mathbf{p}} = \frac12 J_{ud}\left[ 1 + \tau_1 \zeta_1 l_1 \right] + 2J_0 + \Gamma - \frac12\Gamma\left( \gamma_{\bf p}^s + \cos 2\theta \gamma_{\bf p}^d \right) \zeta_1 + \left[2J_0 \gamma_{\bf p}^s + \frac12 \Gamma \left(\gamma_{\bf p}^s + \cos 2\theta \gamma_{\bf p}^d\right)\right]\tau_1 \zeta_1 + D \gamma_{\bf p}^{p} \tau_2 \zeta_2 l_3.
\end{equation}
\end{widetext}
In addition to the $s$-wave harmonic, we  now must introduce the additional cubic harmonics
\begin{subequations}
\begin{align}
    & \gamma_{\bf p}^{p } = \sin \theta \sin p_x - \cos \theta \sin p_y \\
    & \gamma_{\bf p}^d = \frac12 \left[ \cos p_x - \cos p_y \right] ,
\end{align}
\end{subequations}
which depend on the N\'eel order orientation $\theta$ and are of $p$- and $d$-wave symmetries, respectively.
We also have the inter-layer exchange constant $J_{ud}>0$ and intra-layer exchange energies, which are given in Eq.~\eqref{eqn:rashba-exchange-constants}.  

This system has four different magnon bands, which can be split into two different representations based on their eigenvalue under the parity operator $\Pi = \zeta_1l_1$, which commutes with the spin-wave Hamiltonian and assumes the two eigenvalues $\Pi = \pm 1$.
Each of these representations is itself doubly-degenerate.
The dispersions may be obtained analytically, as is done in Ref.~\cite{Bonesteel.1993}, and are given in full in Appendix~\ref{app:bilayer}.
In particular, we find that the even-parity representation ($\Pi = +1$) contains the gapless acoustic pseudo-Goldstone mode, and one gapped optical mode with resonance frequency $\Omega_{0,+-} = \sqrt{ (2J_0 + \Gamma)(2\Gamma + 2J_{ud})}$.
The odd-parity ($\Pi = -1$) representation contains two gapped modes with gaps $\Omega_{0,-+} = \sqrt{J_{ud}( 4J_0 + \Gamma)}$ and $\Omega_{0,--} = \sqrt{ \Gamma( 4J_0 + J_{ud} + \Gamma)}$.

Unless $J_{ud}$ is above a critical value (which is not very large for realistic parameters), it will turn out that this mean-field is dynamically unstable since the acoustic mode energies will become complex, signaling the failure of the ansatz of easy-plane N\'eel-order.
For $\theta = 0$ (N\'eel order along the $a$ axis), we find a critical threshold of order $J_{ud}/J_0 \gtrsim 6 (\lambda /t)^2$, with $\lambda /t \sim .025$ for realistic parameters, which leads to an easily satisfied stability criterion. 
For the parameters of $t=\SI{400}{\meV}$ and $\lambda = \SI{10}{\meV}$, this is satisfied by $J_{ud} \sim \SI{.5}{\meV}$, which we take in the calculations henceforth. 
The equilibrium spin-wave dispersion relation is shown in Fig.~\ref{fig:EP-eq-dispersion}, and is reasonably well described by a simple easy-plane model with additional interlayer coupling.

\begin{figure}
    \centering
    \includegraphics[width=\linewidth]{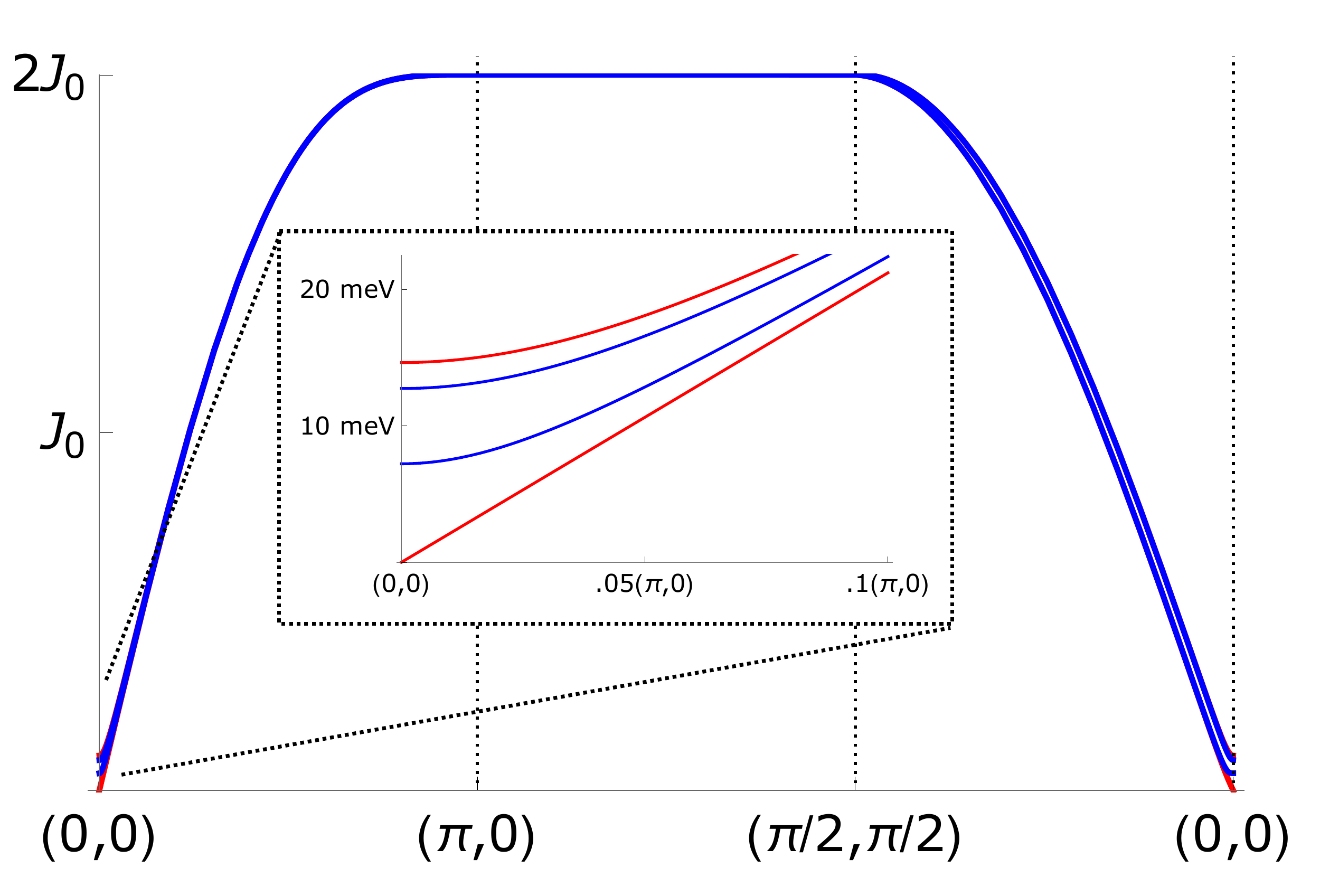}
    \caption{Spin wave spectrum of the bilayer system for realistic parameters of $J_0 =\SI{150}{\meV}, \lambda/t = .017, J_{ud}/J_0 = .0018$ for the case of N\'eel order along the $a$-axis ($\theta = 0$).
    We see that is very close to a doubling of the easy-plane dispersion, except for one of the Goldstone modes is rendered massive by the bilayer coupling.
    (Inset) Expanded view of the behavior near the origin, illustrating the masses of various modes.
    The bands are colored according to their parity under $\Pi = \zeta_1 l_1$ (blue for odd parity, red for even).}
    \label{fig:bilayer-eq}
\end{figure}

Having developed this picture of the equilibrium properties, we now turn our attention towards understanding the role of the magnetoelectric coupling when $g > 0 $. 
We will study this in two parts; first we will study the linear coupling of $E^z$ to the magnons, which produces magnon-polaritons.
We will then examine the quadratic coupling which allows for the squeezing of magnons by the electric field.

\section{Magnon-Polaritons\label{sec:polaritons}}
We now ``switch on" the magnetoelectric coupling $g$. 
We first note that, except for the presence of the layer-index quantum number in the case of the bilayer model, the coupling to in-plane N\'eel order is the same for the two models.
In particular, we will find that there is a linear coupling between the electric field $\hat{E}^z_{\bf q}$ and the spin-wave operators $a_{\bf q},b_{-\bf q}$ which goes as $|\mathbf{q}|$ at long wavelengths,. 
At finite momentum this will produce hybridization between the magnons and photons at finite momentum\textemdash magnon-polaritons. 
In addition to this linear coupling, we will find a term which couples the electric field $\hat{E}^z_{\bf q}$ to magnon bilinears and tends to a constant as the electric field momentum ${\bf q}\to 0$. 
We will investigate this coupling in the next section.
For now, we start by analyzing the easy-plane toy-model.

\subsection{Easy-Plane Model} \label{sec:mono-soc}
We supplement the easy-plane Hamiltonian~\eqref{eqn:ep-ham} with the magnetoelectric interaction term 
\begin{equation}
    \label{eqn:ep-int-ham}
    \hat{H}_{\rm int-EP} = g\sum_{j\in A,\bm \delta} \hat{E}^z_{j,j+\bm \delta}(\mathbf{e}_z\times \bm \delta) \cdot \mathbf{\hat{S}}_{j} \times \mathbf{\hat{S}}_{j+\bm \delta}.
\end{equation}
We now expand around the ground state found in Sec.~\ref{sub:ep-eq} and the details of this procedure are provided in Appendix~\ref{app:EP}.
To first order we find the magneto-electric coupling
\begin{equation}
\label{eqn:ep-sw-int-lin}
\hat{H}_{\rm int}^{(1)}  = \sum_{\bf q} i\frac14 g \hat{E}^z_{-\bf q} \gamma_{\bf q}^{\parallel} \left( a_{\bf q} - b_{\bf q} \right) + \textrm{h.c.} 
\end{equation}
with the new ``longitudinal" form factor 
\begin{equation}
    \gamma_{\bf q}^{\parallel} = 2\cos \theta \sin\frac{q_x}{2} + 2\sin \theta \sin \frac{q_y}{2} \sim q_x \cos \theta  + q_y \sin \theta
\end{equation}
appearing alongside the old ``p-wave" form factor.
The last relation $\sin q_j \sim q_j$ is valid due to the long-wavelengths of the probe electric field, which is an excellent approximation even in the near-field of a Terahertz resonator. 

We now analyze the effect of the linear coupling by determining the linear response of the magnon system to the electric field~\cite{Moriya.1968,Tanabe.1965}. 
Standard methods yield the contribution to the $zz$ element of the dielectric susceptibilty tensor~\footnote{This definition of $\chi^{\rm diel}$ is according to the notation used in, e.g. Jackson, and is not to be confused with the response function $\chi$ as defined in, e.g. Nozieres and Pines, which is actually the density-density response function.} due to the linear coupling (in lattice units) as 
\begin{equation}
    \label{eqn:EP-sus-1} 
    \chi_{zz}^{\rm diel}(\omega,\mathbf{q}) = - \frac{g^2 | \gamma^{\parallel}_{\bf q}|^2 }{4} \frac{2J_0(1+ \gamma_{\bf q}^s ) }{ (\omega+ i0^+)^2 - \Omega_{\bf q,-}^2} .
\end{equation}
The last equality holds for long wavelengths and low frequencies.
We can note a few salient features. 
The first is that the electric field only linearly couples to the massive mode, characterized by $\zeta = -1$, with frequency given by Eq.~\eqref{eqn:EP-dispersion}.
At long wavelengths, this recovers the spin-wave gap $\Delta_{\rm sw} = \sqrt{8J_0\Gamma}$.

The next salient feature is that the oscillator strength is clearly proportional to $\mathbf{q}$ through the longitudinal form factor $\gamma_{\bf q}^\parallel = (\mathbf{N}\cdot\mathbf{q})$, where $\mathbf{N}$ is the direction of the N\'eel vector.
Consequently, this response must arise beyond the dipole approximation, and in general will require near-field techniques to detect~\cite{Sun.20205op,Zhang.2018jbq,Fei.2011,Jiang.2016,Lu.2020,McLeod.2014,Basov.2016}.
This can be anticipated on symmetry grounds by observing that a fluctuation in the magnetization can only couple to the electric field through a matrix element which is odd under spatial inversion.

In the presence of this coupling, the dielectric function becomes 
\begin{equation}
    \label{eqn:EP-diel-lin} 
    \epsilon_{zz}^{(1)}(\omega,\mathbf{q})/\epsilon_{\infty} = 1 - \frac{ F | \gamma^{\parallel}_{\bf q}|^2 (1+ \gamma_{\bf q}^s) }{ (\omega+ i0^+)^2 - \Omega_{\bf q, -}^2} .
\end{equation}
The zeros of this indicate the dispersion of the longitudinal collective modes, which are dubbed ``longitudinal magnon-polaritons."
The oscillator strength is given by 
\begin{equation}
\label{eqn:ep-osc}
    F = \frac{g^2 J_0 }{\epsilon_\infty \mathcal{V} }.
\end{equation}
Here we have replaced the lattice units, with $\mathcal{V}$ the unit-cell volume and $\epsilon_{\infty}$ is the high-frequency bare dielectric constant.
We assume a phonon oscillator strength of $Z^2_{\rm ph}  e^2 / M_{\rm ph} \epsilon_{\rm \infty} \mathcal{V} = S_{\rm ph} \sim 2.8 \Omega_{\rm ph}^2$~\cite{Henn.1997}, which yields an estimate of $F \sim 6.2 \times 10^{4} \si{\meV}^2$.
However, the actual oscillator strength of the resonance is momentum dependent such that it vanishes at long wavelengths, going as $F a^2 \mathbf{q}^2$ (again, replacing lattice units such that $a$ is the ab-plane lattice constant).  
Therefore, at small momentum the splitting off of the longitudinal magnon-polariton collective mode vanishes and coincides with the standard magnon resonance.
In order to couple a cavity to the magnon-polaritons, one has to ensure that the cavity does not simply couple to the $\mathbf{q} = 0$ dielectric response, but also samples features from finite momentum.
We now address this requirement by considering a toy model of a near-field Terahertz resonator.

\begin{figure}
    \centering
    \includegraphics[width=\linewidth]{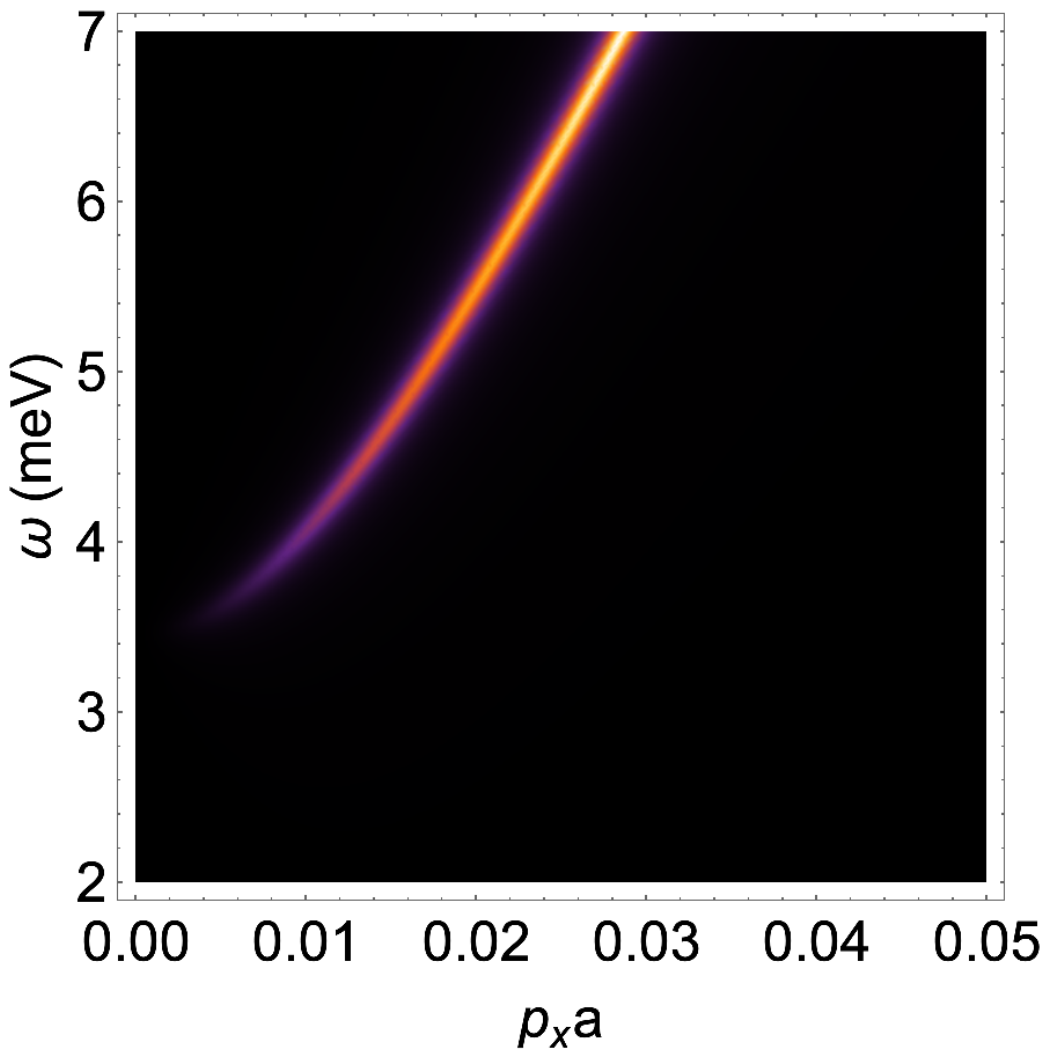}
    \caption{Magnon contribution to the electromagnetic absorption $\Im \chi^{\rm diel}_{zz}(\omega,\mathbf{q})$ for the easy-plane model, from Eq.~\eqref{eqn:EP-diel-lin}. 
    We take characteristic values of $J_0 = \SI{150}{\meV}$, $\Gamma = \SI{.01}{\meV}$, and oscillator strength $F = \SI{6.4E4}{\meV}^2$, and choose consider $\mathbf{q} \parallel \mathbf{\hat{N}}\parallel \mathbf{\hat{e}}_x$.
    We note that the effective oscillator strength vanishes as $\mathbf{q}\to 0$, due to the hybridization form factor $\gamma^{\parallel}_{\bf q}$.
    We also note that the dispersion of the spin wave sets in fairly rapidly. }
    \label{fig:EP-chizz-fig}
\end{figure}

\subsection{Cavity-Magnon Polaritons\label{sub:near-field}}

The combined requirement of a strong Terahertz field and strong Terahertz field-gradient requires us to couple via a near-field coupling scheme~\cite{Benz.2016,Maissen.2019,Bitzer.2009,Zhang-thesis.2019,Lu.2020,Zhang.2018jbq,Fei.2011,McLeod.2014,Cvitkovic.2007,Anderson.2005,Isakov.2017,Benz.2016,Chen.2020,Chen.2014,Chen.2003,Stinson.2014,Sun.20205op,Jiang.2016,Yang.2016}.
There are many routes by which this is achievable; we will only explore one potential technique, which essentially combines a Terahertz resonator~\cite{Maissen.2014ta,Scalari.2012,Yen.2004,Geiser.2012,Zhang.2016nsw,Benea-Chelmus.2020,Duan.2019,Keiser.2021} with a SNOM-type setup in order to generate the near-field coupling~\cite{Benz.2016,Maissen.2019,Bitzer.2009,Zhang-thesis.2019,Lu.2020,Zhang.2018jbq,Fei.2011,McLeod.2014,Cvitkovic.2007,Anderson.2005,Isakov.2017,Benz.2016,Chen.2020,Chen.2014,Chen.2003,Stinson.2014,Sun.20205op,Jiang.2016,Yang.2016}.
This particular model is expanded upon in Appendix~\ref{app:cavity}.
We emphasize that this is only a qualitative discussion of one possible route towards engineering this coupling. 

For our purposes, it will be sufficient to model the resonator as an RLC circuit with linewidth $\kappa \sim R/L$ and resonance $\Omega_{\rm cav}\sim 1/\sqrt{LC}$, where $L,R,C$ are the effective inductance, resistance, and capacitance, respectively~\cite{Bagiante.2015,Wu.2020,Maissen.2014ta,Yen.2004,Scalari.2012,Duan.2019}.
In order to generate the near-field coupling we imagine placing a small metal object, with size of order $R$ within the electric-field antinode of the resonator.
For more details, we refer the reader to Appendix~\ref{app:cavity}, with a schematic depiction shown in Fig. \ref{fig:near-field}.
In the good metal limit, this object will become polarized by the Terahertz electric field, with induced polarization
\begin{equation}
    \mathbf{P}_{\rm ind} = 4\pi \epsilon_0 R^3 \mathbf{E}_{\rm cav}.
\end{equation} 
Thus, the electric field of the cavity $\mathbf{E}_{\rm cav}$ induces an effective dipole which oscillates along with the resonator.
We then assume that the sample has a thickness $h$ and is placed a depth $d\sim R$ below the induced dipole.
In the presence of the induced dipole, the magnons will contribute to the electrostatic energy of the resonator and therefore change the effective capacitance, such that it includes a contribution from the finite-momentum $\chi^{\rm diel}_{zz}(\omega,\mathbf{q})$, where the relevant momenta are of order $|\mathbf{q}| \sim 1/d$.

In Appendix~\ref{app:cavity} we use a simple electrostatic model to evaluate the magnonic substrate contribution to the effective capacitance $K_{\rm magnon}(\omega) = C(\omega)/C_{\rm bare} -1 $. 
We find 
\begin{equation}
\label{eqn:EP-K-func}
    K_{\rm magnon}(\omega) = \frac{4\pi^2 R^6  h}{V_{\rm eff}} \int_{\bf q} \frac{\chi^{\rm diel}_{zz}(\omega,\mathbf{q})}{\epsilon_0} e^{-2d |\mathbf{q}|} |\mathbf{q}\cdot\mathbf{e}_{\rm cav} |^2 .
\end{equation}
Here $\mathbf{e}_{\rm cav}$ is the unit vector corresponding to the polarization of the electric field at the field anti-node, which we take to lie parallel to the $ab$-plane, and $h$ is the sample thickness, which we take in the limit $h \ll d$. 
For the result in the $h \sim d$ limit, we refer the reader to Appendix~\ref{app:cavity}.
We also have introduced the ``effective volume" of the cavity, $V_{\rm eff}$ which is interpreted as the effective volume occupied by the electric field, weighted by the distribution of the electric-field energy density. 
In the parallel plate model, we would find $V_{\rm eff} = A_{\rm eff} l_{\rm eff}$ where $A_{\rm eff}$ is the cross-sectional area of the plates and $l_{\rm eff}$ is their separation. 

We evaluate this contribution, taking oscillator strength estimated in Eq.~\eqref{eqn:ep-osc} of $F \sim 6\times 10^4 \si{\meV}^2$.
The conventional notion of ``strong-coupling" and ``ultra-strong-coupling" do not strictly apply in this sense since the cavity is coupled to a continuum of magnon modes, each with a different resonance and hybridization matrix element.
Nevertheless, we may attempt to introduce a suitable notion of ``hybridization constant," $G_{\rm eff}$ by examining the scaling of the momentum space integral above. 
The cavity electric-field profile has a Fourier transform which has typical momenta of order $q\sim 1/d$.
At this momentum, the effective oscillator strength is of order $F\times (a/d)^2 $ (we replace explicit lattice units here), and the remaining terms in the integrand ultimately scale as $q^4\sim 1/d^4$, such that we arrive at the dimensional-analysis estimate 
\begin{equation}
G_{\rm eff}^2 \sim \frac{ h (R/d)^6 a^2 }{V_{\rm eff}}\times F ,
\end{equation}
as a notion of ``typical" coupling strength, which is to be compared against the magnetic zone-center magnon resonance $\Omega_{\rm sw}$. 

Assuming $R \sim d$, this ends up being suppressed by a factor of roughly $a^2 / A_{\rm eff}$, where $A_{\rm eff}$ is the effective cross-sectional area of the capacitor.
In what follows we take an effective mode volume of $V_{\rm eff} \sim (40\si{\nm})^3$, a sample thickness of $h \sim 10 a$, and take $R =d$ for the purposes of our calculations.
In this regime, we do find the possibility for $G_{\rm eff} \sim \Omega_{\rm sw}$, that is ``strong-coupling."
It is also worth remarking that our toy model is quite crude and a more realistic and optimized model may find even better coupling strengths.

\begin{figure}
    \centering
\includegraphics[width=\linewidth]{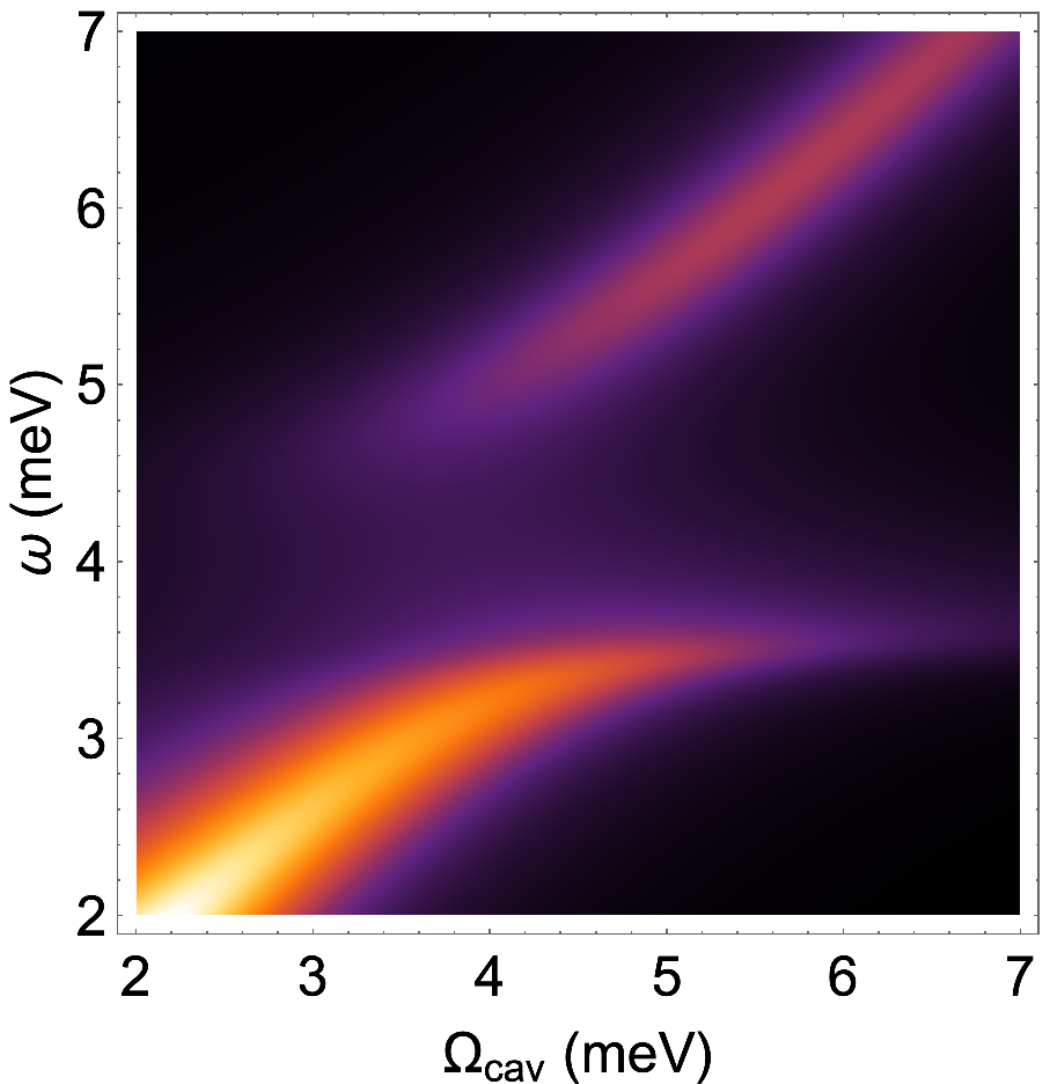}
\caption{Cavity spectral function including capacitive coupling to magnons obtained from Eq.~\eqref{eqn:EP-K-func}.
We take $V_{\rm eff} =(\SI{40}{\nm})^3$, $d\sim R \sim 300 a$, $h \sim 10 a$, and parameters of spin-wave system from main text. 
We assume a cavity linewidth of $\kappa = 1\si{\meV}$.
We see a clear avoided crossing in the cavity spectral function as the cavity comes in to resonance with the spin-wave.  }
\label{fig:EP-avoided-crossing}
\end{figure}

We confirm our dimensional-analysis estimate by numerically evaluating Eq.~\eqref{eqn:EP-K-func} and computing the resulting dressed cavity spectral function $A_{\rm cav}(\omega)$ as a function of the bare cavity resonance.
In the presence of a finite $K_{\rm magnon}(\omega)$, we find the cavity spectral function (including Ohmic damping) of 
\begin{equation}
    A_{\rm cav}(\omega) = -\frac{1}{\pi}\Im \frac{1}{\omega^2 + i\kappa \omega - \Omega_{\rm cav}^2/(1 + K_{\rm cav}(\omega) )},
\end{equation}
as derived in App.~\ref{app:cavity}.
This is illustrated in Fig.~\ref{fig:EP-avoided-crossing}.
We see that as the bare cavity resonance $\Omega_{\rm cav}$ passes through the bulk spin-wave resonance frequency\footnote{Actually it is not exactly equal to the bulk magnon resonance due to contributions from finite momentum magnons. It nevertheless remains close, at least for the parameters chosen here.}, the dressed cavity spectral function experiences level repulsion, indicating our crude estimate of the strong-coupling regime was correct. 

\subsection{Bilayer-Rashba Model} \label{sec:bi-soc}
The bilayer Rashba Hamiltonian is only slightly more complicated to analyze than the easy-plane Hamiltonian.
The bulk of the calculation is carried out in Appendix~\ref{app:bilayer} and we merely present the result here. 
By expanding the interaction Hamiltonian in Eq.~\eqref{eqn:full-me-ham} up to quadratic order we obtain essentially the same linear interaction, but now summed over the two layers
\begin{equation}
\label{eqn:bilayer-sw-int-lin}
H_{\rm int}^{(1)} =  \frac{ig}{4}\sum_{\bf p \ell}\hat{E}^z_{-\bf p} \gamma_{\bf p}^{\parallel} \left( a_{\bf q\ell} - b_{\bf q\ell} \right)+\textrm{h.c.} .
\end{equation}

We apply the same methods as in the easy-plane case, however we must now also keep track of the layer quantum number. 
We see immediately that the electric field only couples to the $\zeta_1 = -1$ and $l_1 = +1$ modes. 
The complication is that, in the presence of the DM interaction, $l_1$ is no longer a good quantum number.
Therefore, we evaluate in the energy eigenbasis and then take the appropriate matrix element, producing the dielectric susceptibility 
\begin{equation}
    \label{eqn:rashba-sus-lin}
    \chi^{\rm diel}_{zz}(\omega,\mathbf{q} ) = - \frac{g^2 |\gamma^{\parallel}_{\bf q}|^2}{16} w^T \cdot \hat{\mathbb{D}}^R(\omega,\mathbf{p})\cdot w,
\end{equation}
where $w^T =(1,1,-1,-1,1,1,-1,-1)$ is the (non-normalized) Nambu spinor corresponding to the direct product of the $+1$ eigenstates of $\tau_1$ and $l_1$,  and $-1$ eigenstate of $\zeta_1$, and $\mathbb{D}^R =\left( (\omega + i0^+)\tau_3 - \mathbb{\hat{M}}_{\mathbf{p}} \right)^{-1}$ is the retarded spin-wave propagator for the bilayer system, with $\mathbb{\hat{M}}_{\mathbf{p}}$ the quadratic form from Eq.~\eqref{eqn:eq-rashba-spin-wave}.

\begin{figure*}
    \centering
    \includegraphics[width=\linewidth]{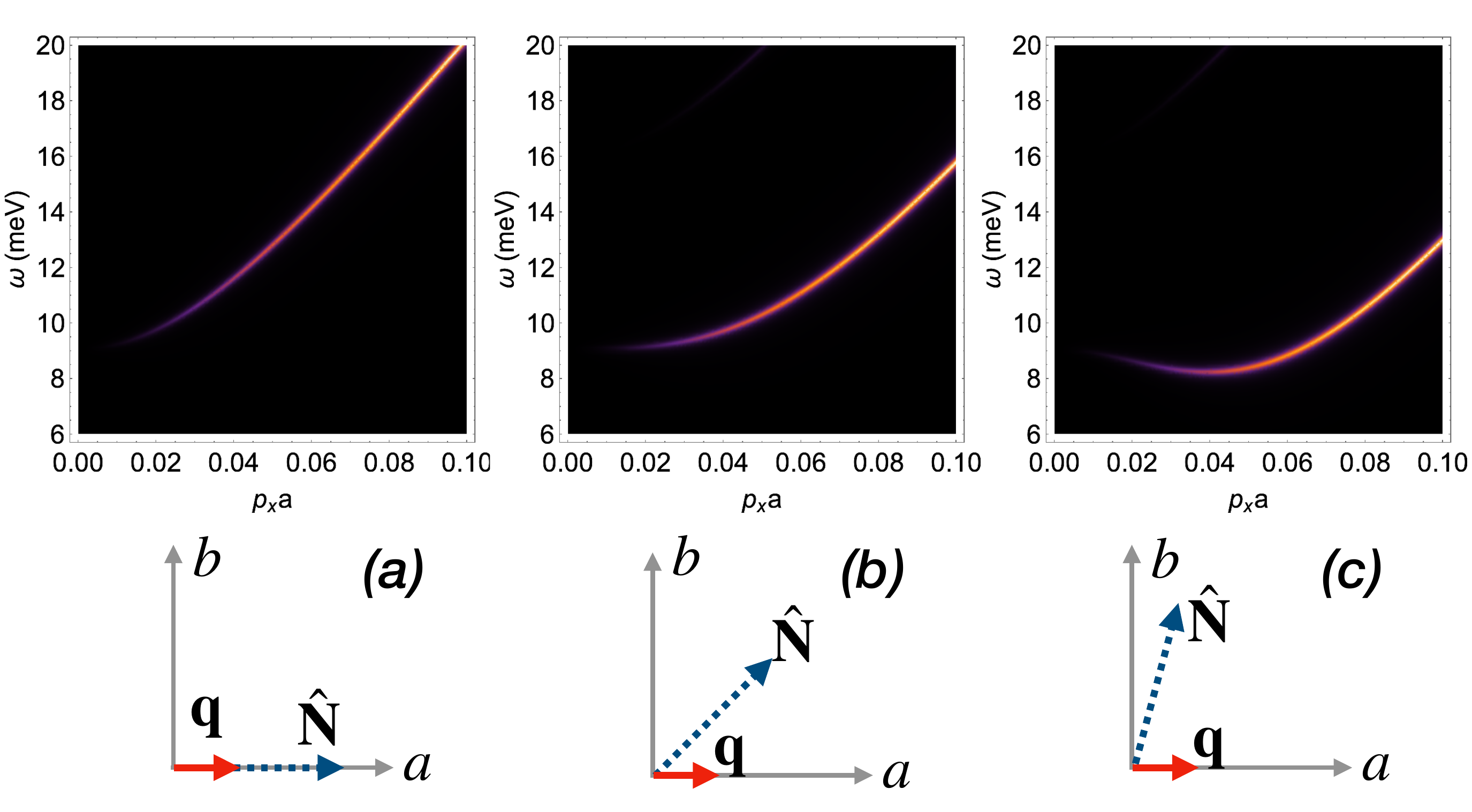}
    \caption{Magnon contribution to the $\Im \chi^{\rm diel}_{zz}(\omega,\mathbf{q})$ for the Rashba bilayer model, from Eq.~\eqref{eqn:rashba-sus-lin}.
    We again observe the vanishing oscillator strength at long-wavelengths.
    In this case, the dispersion relation of the magnon is less trivial and we illustrate the susceptibility for three different configurations of the the in-plane momentum and Neel order, illustrated in the lower part of the figure.
    This is shown specifically for three scenarios of Neel order at angles relative to the $a$-axis of (a) $\theta = 0$, (b) $\theta = 45^\circ$, and (c) $\theta = 75^\circ$ respectively. 
    We for simplicity only show $\mathbf{q}\parallel a$.
    While it is not visible on the chosen scale, there is a small contribution for a higher-lying spin-wave band which behaves similarly as the contribution from the dominant band.  }
    \label{fig:BL-chizz-fig}
\end{figure*}

The form of the dissipative part of the dielectric susceptibility in Eq.~\eqref{eqn:rashba-sus-lin} is depicted in Fig.~\ref{fig:BL-chizz-fig} for a few different orientations of the in-plane Neel vector relative to the $a$-axis, while scanning momentum $\mathbf{q}$ also along the $a$-axis.
We note this has a slightly more complicated behavior due to the absence of spin-rotation symmetry, but broadly speaking it is qualitatively similar to the easy-plane system.
What is not visible in the spectral plots of Fig.~\ref{fig:BL-chizz-fig} is that another higher-lying magnon branch also is optically active at finite momentum, but has a much smaller oscillator strength and we will not dwell on this mode.

\begin{figure}
    \centering
\includegraphics[width=\linewidth]{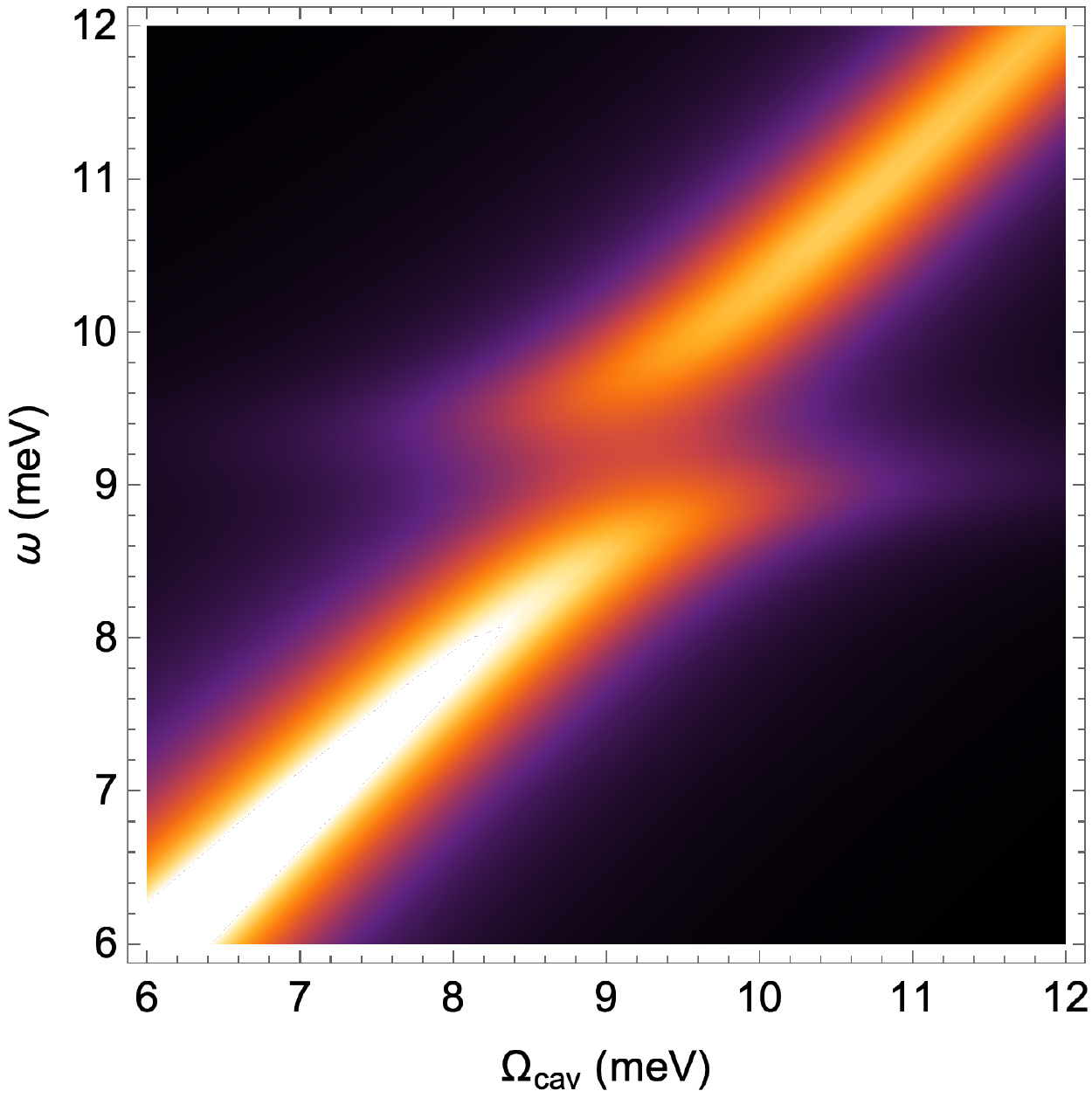}
\caption{Cavity spectral function including capacitive coupling to magnons obtained from Eq.~\eqref{eqn:EP-K-func} and using $\chi^{\rm diel}_{zz}$ as obtained in Eq.~\eqref{eqn:rashba-sus-lin}.
We take $V_{\rm eff} =(\SI{40}{\nm})^3$, $d\sim R \sim 300 a$, $h \sim 10 a$, and parameters of spin-wave system from main text. 
We assume a cavity linewidth of $\kappa = 1\si{\meV}$.
We see a clear avoided crossing in the cavity spectral function as the cavity comes in to resonance with the spin-wave.  }
\label{fig:BL-avoided-crossing}
\end{figure}

We again analyze the near-field interaction with a Terahertz cavity using the setup outline in the previous subsection. 
The derivation of Eq.~\eqref{eqn:EP-K-func} remains valid, if we use the correct dielectric susceptibility $\chi^{\rm diel}_{zz}$, which was just computed above in Eq.~\eqref{eqn:rashba-sus-lin}.
Likewise, the cavity spectral function retains the same dependence on $K_{\rm magnon}$.
This is depicted for a particular orientation of Neel order in Fig.~\ref{fig:BL-avoided-crossing}.

As we can see, the result is qualitatively similar to the case of the single monolayer.
To summarize, we clearly see that at finite momentum spin-orbit coupling admixes spin-waves and optical phonons, such that magnons acquire a finite electric dipole moment.
This then allows for their coupling to cavity photons, provided there is a sufficiently large electric field gradient.
In the section above, we outline one possible way to generate the necessary field gradient\textemdash however, we emphasize that the scheme proposed above is only one of many potential ways this may be done.
Other promising routes may involve making smaller split-ring resonators with larger fringing fields, using wave-guide based modes~\cite{Mandal.2020,Ashida.2021}, plasmonic nanocavities~\cite{Benz.2016}, or employing photonic crystal metamaterials~\cite{Sunku.2018}.   

In contrast, in the next section we will consider a different coupling mechanism which allows for the cavity to strongly couple to pairs of magnons without the need to introduce near-field coupling schemes.

\section{Bimagnon Interaction\label{sec:bimagnon}}
In this section we will demonstrate how, in a bilayer system like YBCO, the same optical phonon can also lead to a coupling between the cavity electromagnetic field and pairs of magnons\textemdash bimagnons. 
Furthermore, this mechanism doesn't rely on spin-orbit coupling and therefore is anticipated to be much stronger. 
In principle this mechanism can be used to generate correlated pairs of magnons~\cite{Juraschek.2020,Michael.2020,Rajasekaran.2016,Dolgirev.2021,Malz.2019,Bukov.2015} via parametric driving from the cavity field, but in this work we will focus on the linear response regime and leave a more in-depth analysis to future work. 

For brevity, we will focus on the linear response of the bilayer system to a homogeneous electric field and limit our attention to the larger $g'$ coupling.
As a reminder, this coupling is due to the buckled nature of the bilayer structure in equilibrium, and is independent of spin-orbit coupling.
In the buckled structure, the Cu-O-Cu bond length and angle vary linearly with phonon displacement, and oppositely between the two layers (such that overall inversion symmetry is preserved).

The relevant term in the Hamiltonian is 
\begin{equation}
\label{eqn:bilayer-int}
    H_{\rm int} = \sum_{j\in A\bm\delta\ell} g'(-1)^\ell \hat{E}^z \mathbf{S}_{j\ell}\cdot\mathbf{S}_{j+\bm \delta \ell},
\end{equation}
where $g' = Z_{\rm ph} e/ M_{\rm ph} \Omega_{\rm ph} \times 8 t \alpha/ U$, and $\alpha = \frac{dt}{dQ}|_{\rm eq}$ is the equilibrium linear variation of the nearest-neighbor hopping with respect to the oxygen displacement. 
We see the $(-1)^\ell$ factor ensures the coupling has odd parity under inversion, which swaps the two layers and also inverts the electric field z-component $\hat{E}_z$.

Applying the linear spin-wave expansion, we find that the lowest order coupling due to Hamiltonian~\eqref{eqn:bilayer-int} is quadratic in the Holstein-Primakoff bosons. 
We obtain the coupling (at zero electric field momentum)
\begin{equation}
\label{eqn:bl-quad-int}
\hat{H}_{\rm int}^{(2)} = \frac12 \sum_{\bf p} \Psi_{\bf p}^\dagger \hat{E}^z \mathbb{\hat{V}}_{\bf p}\Psi_{\bf p},
\end{equation}
with the scattering vertex 
\begin{equation}
\label{eqn:bl-quad-vertex}
\mathbb{\hat{V}}_{\bf p} =  2g' l_3 \left[ 1 +\gamma_{\bf p}^s \tau_1 \zeta_1 \right].
\end{equation}
Here $\Psi_{\bf p}$ is the eight-component Nambu spinor from Section~\ref{sec:eq-spin-waves}.

\begin{figure}
    \centering
    \includegraphics[width=\linewidth]{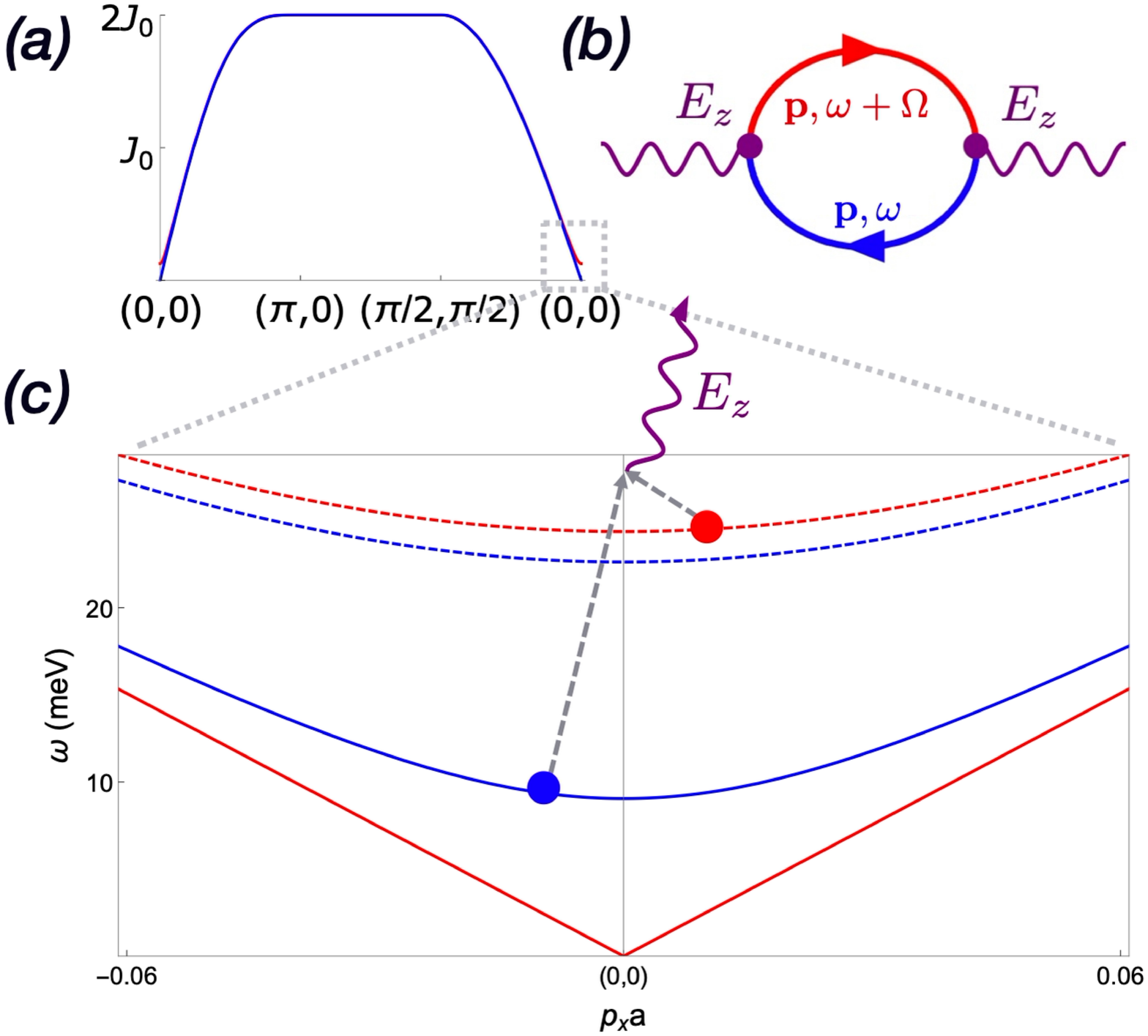}
    \caption{
    Magnon bands in the bilayer system: (a) The set of four magnon bands for $J_0 = 128$ \si{meV}, $J_{ud} = 0.5$ \si{meV}, and $\Gamma = 0.016$ \si{meV}; (b) Feynman diagram illustrating the bimagnon contribution to the dielectric susceptibility; (c) Expanded view of the magnon band dispersion near the center of the Brillouin zone, with the interband pairs of magnons with opposite momenta shown schematically. }
    \label{fig:bilayer-pair}
\end{figure}

In order to simplify the calculations, we will replace the full microscopic Rashba bilayer model with an approximately equivalent model of two layers with in-plane easy-plane anisotropy and weak interlayer coupling, such that the relevant quadratic form for spin-waves is  $\mathbb{\hat{M}}_{\bf p} =  \frac12 J_{ud}\left[ 1 + \tau_1 \zeta_1 l_1 \right] + 2J_0 - \Gamma \gamma_{\bf p}^s \zeta_1 + (2J_0 - \Gamma)\gamma_{\bf p}^s \zeta_1\tau_1 $.
This dispersion is depicted in Fig.~\ref{fig:bilayer-pair}(a), and in more detail near the $\Gamma$-point in Fig.~\ref{fig:bilayer-pair}(c).
We find, in general, four bands with dispersions
\begin{equation}
\label{eqn:bl-scalar-dispersions}
    \Omega_{{\bf p},l}^{(\zeta)} = \sqrt{ \left( 2J_0 - \Gamma \gamma_{\bf p}^s \zeta + \frac12 J_{ud} \right)^2 - \left( \left[2J_0 - \Gamma \right]\gamma_{\bf p}^s + \frac12 J_{ud} l \right)^2},
\end{equation}
where $\zeta = \pm$ is the parity under sublattice symmetry $\zeta_1$, and $l=\pm 1$ is the parity under bilayer-inversion $l_1$.

In order to compute the response, we use the Matsubara finite-temperature method to compute the magnon contribution to the electromagnetic response function, corresponding to the Feynamn diagram in Fig.~\ref{fig:bilayer-pair}(b).
We present the result at $T = 0$, and after performing the relevant analytic continuations.

Since the scattering vertex in Eq.~\eqref{eqn:bl-quad-vertex} commutes with the sublattice parity $\zeta_1$, we can express the resulting response function $\chi^{\rm diel}_{zz}$ as a sum over two processes\textemdash one for each value of $\zeta$.
Furthermore, because the vertex flips the interlayer parity $l_1$, we find that only interband processes contribute to $\chi^{\rm diel}$, as depicted in Fig.~\ref{fig:bilayer-pair}(c).
We therefore find 
\begin{equation}
    \chi^{\rm diel}_{zz}(\omega) = \chi^{(+)}_{zz}(\omega) + \chi^{(-)}_{zz}(\omega),
\end{equation}
indicating the contribution from the bands with $\zeta = +1,-1$ respectively.
The full result of this calculation is cumbersome and relegated to Appendix~\ref{subsec:Phonon-magnon-coupling}.
In the following, we will numerically evaluate the resulting expressions, as well as prevent simplified results in the appropriate regimes.

We begin by studying the limiting case when anisotropy $\Gamma = 0$.
At this point the result simplifies further since both of the contributions $\chi^{(\pm)}$ become degenerate. 
This leads to the analytic form  
\begin{equation}
\chi_{zz}^{\rm diel}\left(\omega\right)=\int_{{\bf p}}\frac{8(g')^2[1-(\gamma_{{\bf p}}^{s})^{2}]\left(\Omega_{-,{\bf {p}}}-\Omega_{+,{\bf {p}}}\right){}^{2}\left(\Omega_{-,{\bf {p}}}+\Omega_{+,{\bf {p}}}\right)}{\Omega_{-,{\bf {p}}}\Omega_{+,{\bf {p}}}\left[\left(\Omega_{-,{\bf {p}}}+\Omega_{+,{\bf {p}}}\right){}^{2}-\left(\omega+i0^{+}\right)^{2}\right]}.\label{eqn:BL-scalar-chi}
\end{equation}
Here we have suppressed the dependence on the $\zeta$ quantum number, since the $\zeta = \pm 1$ bands are both degenerate in this limit.

\begin{figure}
\begin{centering}
\includegraphics[width=\linewidth]{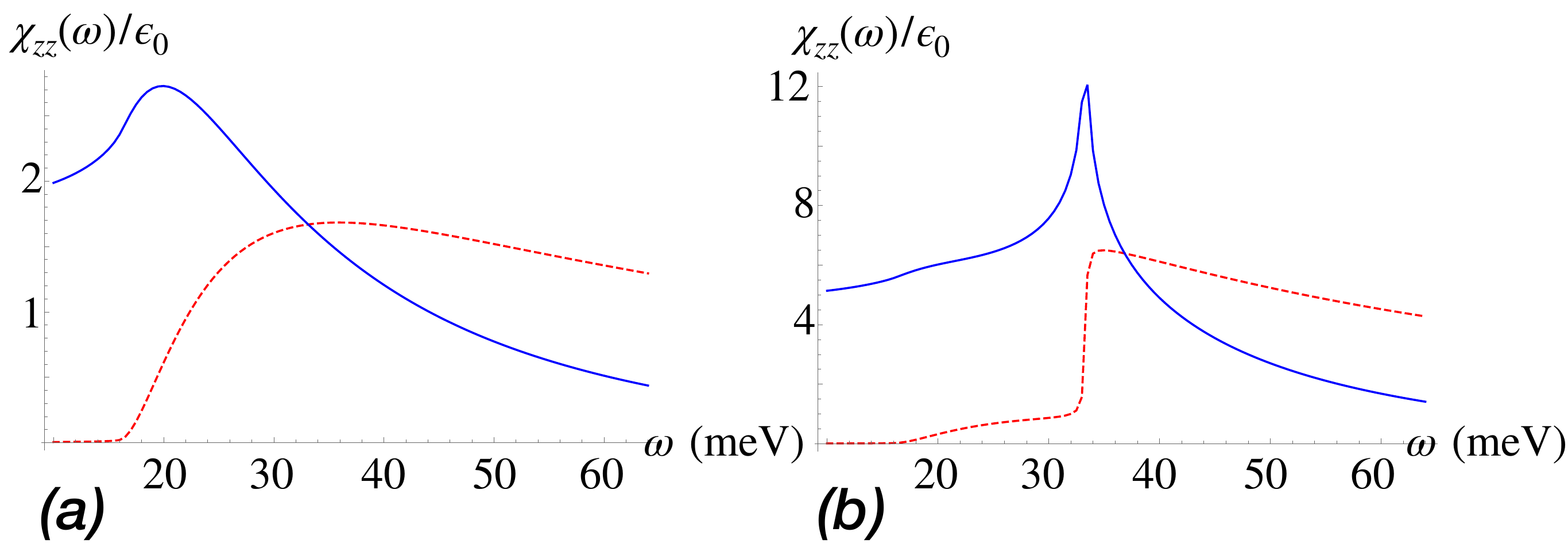} 
\par\end{centering}
\caption{Contribution to the dielectric susceptibility due to bimagnon coupling, corresponding to the Feynman diagram in the inset of Fig.~\ref{fig:bilayer-pair}(b). 
We show the real part (solid blue), and imaginary part (dashed red) of the relative susceptibility $\chi^{\rm diel}_{zz}(\omega,\mathbf{q} = 0)/\epsilon_0$ as a function of the electric field frequency $\omega$.
We use scalar exchange $J_0 = \SI{128}{\meV}$ and take interlayer exchange $J_{ud} = \SI{.5}{\meV}$, compatible with our estimates from Sec.~\ref{sub:rashba-eq}.
(a) Response when easy-plane anisotropy =0, corresponding to the case of two doubly-degenerate magnon bands.
The response sets in once the single-magnon threshold is passed since in this case the bimagnon is actually a single gapped magnon, along with a soft Goldstone mode.
(b) Response when $\Gamma = \SI{.16}{\meV}$, the value obtained in previous sections. 
The dominant feature corresponds to a bimagnon excitation gap $\Omega_{{\rm bimag}}\sim \SI{33}{\meV}$. 
In contrast to (a), we find a strong response at this frequency driven by the Van Hove singularity, which in two dimensions is logarithmic for the real-part and step-like for the imaginary part.
}
\label{fig:BL-scalar-chi} 
\end{figure}

There is a non-trivial response even when $\Gamma = 0$, with the corresponding dielectric susceptibility $\chi^{\rm diel}_{zz}$ shown in Fig.~\ref{fig:BL-scalar-chi}(a). 
We see that the response is fairly muted, and mostly corresponds to an onset in the electromagnetic absorption once the threshold for magnon pair-creation is surpassed.
Note that because when $\Gamma = 0$, one magnon branch involved in the pair processes is gapless, such that the two-magnon threshold energy coincides with the single-magnon threshold energy of the gapped band. 

In contrast, when $\Gamma > 0$ we find the response exhibits a singular response due to the Van Hove singularity in two-dimensions.
This is illustrated in Fig.~\ref{fig:BL-scalar-chi}(b).
We see that there is a meager response as the frequency passes through the single-magnon threshold, but this is largely overshadowed by the response at the two-magnon threshold corresponding to the production of two gapped magnons, depicted in Fig.~\ref{fig:bilayer-pair}(c) and also clearly visible in the dielectric response shown in Fig.~\ref{fig:BL-scalar-chi}(b).

To understand the origin of this strong bimagnon response, we study the contribution $\chi^{(-)}_{zz}(\omega)$, which originates from the production of two magnons, one on each of the the $\zeta = -1$ bands.
We find this assumes the form 
\begin{equation}
\label{eqn:bl-scalar-van-hove}
    \chi_{zz}^{\left(-\right)}\left(\omega\right)= \int_{\bf p }\frac{ C_{\bf p}^{(-)} }{\left(\Omega_{\rm p-}^{(-)}+\Omega_{{\rm p}+}^{(-)}\right)^2-(\omega+i0^{+})^{2}},
\end{equation}
where we recall that $\Omega_{{\bf p}, l}^{(\zeta)}$ are given in Eq.~\eqref{eqn:bl-scalar-dispersions} and $C_{\bf p}$ is a complicated expression involving matrix elements, which importantly tends to a constant as $\mathbf{p} \to 0$.
Therefore, at low frequencies we may expand the integrand for small momentum $\mathbf{p}$ and invoke rotational symmetry of the resulting expansion, such that the dispersion may be characterized by the bimagnon dispersion 
\begin{equation}
\Omega_{\rm p-}^{(-)}+\Omega_{{\rm p}+}^{(-)}  \sim \Omega_{\rm bimag}+ \frac{1}{2M_{\rm bimag}}\mathbf{p}^2 + O(|\mathbf{p}|^3 ) ,
\end{equation}
where the bimagnon resonance occurs at 
\begin{equation}
\Omega_{\rm bimag} =  \Omega_{-,0}^{(-)} + \Omega_{+,0}^{(-)} = \sqrt{8\Gamma J_0}+\sqrt{8J_0(\Gamma+J_{\text{ud}}/2)}.
\end{equation}
We now see that the resulting integral acquires a singular contribution due to the quadratic nature of the dispersion of the bimagnon state, which leads to a logarithmically-divergent contribution to the real-part of the response function $\chi_{zz}^{(-)}$ in two-dimensions. 

In contrast, this is not the case for the other contribution since, as previously established, the bimagnon excitation involves one gapped spin-wave and one soft Goldstone mode which therefore yields a linear dispersion at sufficiently small momenta.
We will however note in passing that in principle due to the lower orthorhombic symmetry of real cuprate materials, there are further magnetocrystalline anisotropies which will ultimately gap out this Goldstone mode, perhaps leading to another singular contribution. 
We leave the study of these more realistic models to a future work~\footnote{In principle, one should also include other effects such as those implied by the full Rashba bilayer model, spin-wave broadening due to interactions or impurities, or the effects of weak $c$-axis exchange. The fate of the Van Hove singularity upon including all of these effects is clearly more complicated, albeit interesting and relevant to future studies.}.

\begin{figure}
\begin{centering}
\includegraphics[width=\linewidth]{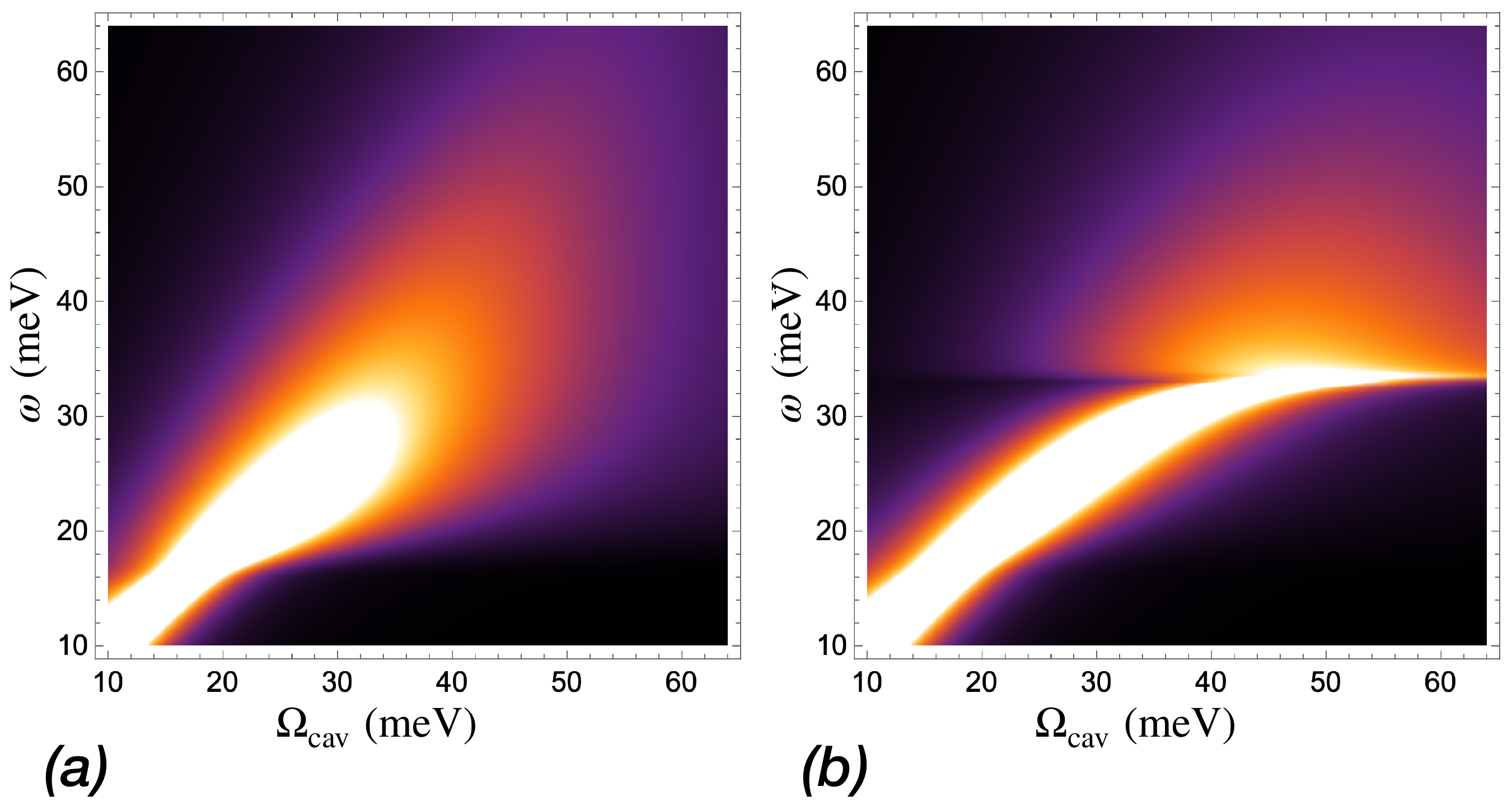} 
\par\end{centering}
\caption{Cavity spectral functions include contribution to the dielectric susceptibility due to bimagnon coupling illustrated in Fig.~\ref{fig:BL-chizz-fig}.
(a) Density plot illustrating the behavior of the cavity spectral function $A_{\rm cav}(\omega)$ as the bare cavity resonance frequency $\Omega_{\rm cav}$ is tuned across the bimagnon resonance for the case of $\Gamma = 0$, which corresponds to $\chi_{zz}$ from Fig.~\ref{fig:BL-chizz-fig}(a). 
Upon crossing the threshold for pair production, the imaginary part of $\chi^{\rm diel}$ sets in rapidly, leading to a mostly dissipative response and resulting diffuse line shape. 
(b) Density plot for cavity spectral function with finite $\Gamma = \SI{.16}{\meV}$.
At the edge of the bimagnon continuum, located around $\omega = \SI{33}{\meV}$, the logarithmic singularity manifests as a visible feature in the cavity spectral function.
Though not a true avoided crossing due to the onset of strong damping by the continuum, we do see some type of level repulsion as we attempt to tune the bare cavity frequency $\Omega_{\rm cav}$ through the resonance.
Note the cavity spectral function is renormalized such that the resonance occurs at $\omega^2 = \Omega_{\rm cav}^2$ as $\Omega_{\rm cav}^2 \to 0$.
}
\label{fig:BL-scalar-cav} 
\end{figure}

Finally, given that we find a strong bimagnon contribution to the dielectric susceptibility, we consider a simple model of coupling to a cavity~\cite{Maissen.2014ta,Scalari.2012,Bagiante.2015,Yen.2004,Wu.2020,Duan.2019}, in order to assess whether we may find resulting ``bimagnon-polaritons."
Using the computed dielectric response functions $\chi^{\rm diel}_{zz}(\omega)$, we determine the effective capacitance of a hypothetical cavity enclosing the cuprate sample, with homogeneous field polarized along the $c$-axis, such that $C(\omega) = A_{\rm eff}/\ell_{\rm eff} \epsilon_{zz}(\omega)$.
We then study the cavity spectral function 
\begin{equation}
A_{\rm cav}(\omega) = -\frac{1}{\pi} \Im \frac{1}{\omega^2 + i\gamma \omega - \Omega_{\rm cav}^2 \epsilon_{zz}(0)/\epsilon_{zz}(\omega) }.
\end{equation}
The factor of $\epsilon_{zz}(0)/\epsilon_{zz}(\omega)$ arises because we have specifically chosen to parameterize $\Omega_{\rm cav}$ such that the dressed resonance agrees with the bare cavity frequency $\Omega_{\rm cav} $ as $\Omega_{\rm cav}\to 0$.
The resulting cavity spectral function is shown in Fig.~\ref{fig:BL-scalar-cav}(a) for the case of $\Gamma = 0$, and Fig.~\ref{fig:BL-scalar-cav}(b) for $\Gamma >0$ (parameter choice in caption).

We see that neither case exhibits a conventional strong-coupling splitting near the bimagnon feature, due to it being more like a two-particle continuum than true resonance.
In the case of $\Gamma =0 $, we see that once the cavity resonance crosses the peak in $\Re \epsilon(\omega)$, it rapidly dissolves into a diffuse line shape.
While this does not mean that the coupling is weak (in fact, it implies the coupling is strong enough to substantially damp the cavity), it does mean that the resonance is too broad to participate in coherent level repulsion and is largely dissipative in nature. 

\begin{figure}
\begin{centering}
\includegraphics[width=\linewidth]{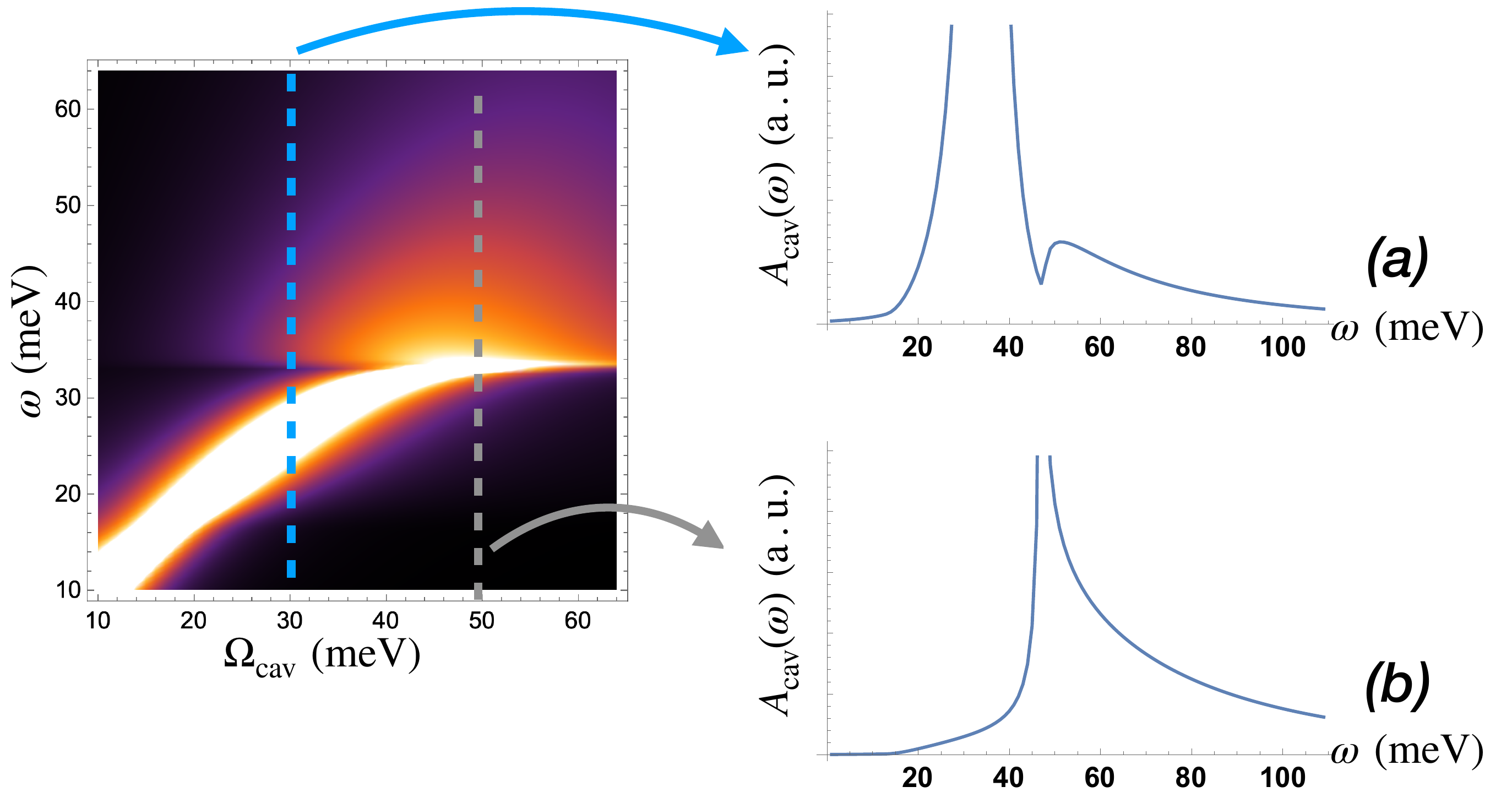} 
\par\end{centering}
\caption{Line cuts of cavity spectral function from Fig.~\ref{fig:BL-scalar-cav}(b), with finite $\Gamma = \SI{.16}{\meV}$ for constant $\Omega_{\rm cav}$.
(a) Line cut along $\Omega_{\rm cav} = \SI{30}{\meV}$. 
We see a small secondary shoulder-like feature appear in the cavity spectral function at the bimagnon resonance, but it is largely overshadowed by the bare resonance of the cavity.
(b) As the cavity is tuned through the resonance, such that now $\Omega_{\rm cav} = \SI{50}{\meV}$, the shoulder-like feature becomes washed out be a broad overdamped continuum.
In this case, the dominant effect of the magnons is to provide an asymmetric damping to the cavity lineshape, which essentially just reflects this contribution.
}
\label{fig:BL-scalar-line-cut} 
\end{figure}

For finite $\Gamma>0$, shown in Fig.~\ref{fig:BL-scalar-cav}(b), we see a more complicated line-shape emerge due to the singular nature of the bimagnon coupling. 
We investigate this further in Fig.~\ref{fig:BL-scalar-line-cut} by examining line cuts of the spectral function at constant cavity resonance frequency $\Omega_{\rm cav}$.
We see the strong coupling leads to an asymmetric split-lineshape which shows some signatures of a coherent avoided crossing.  
Increasing the cavity resonance further we find the coherent contribution is quickly overcome by the strong damping, which now dominates the cavity spectral function.
Therefore, it is plausible that in the presence of the strong bimagnon coupling for finite $\Gamma > 0$, strong coherent cavity-magnon interactions may also be observed, provided the cavity lies below the continuum, which rapidly destroys any notion of avoided crossing.

\section{Conclusion\label{sec:conclusion}}
In conclusion, we have examined a variety of ways in which a Terahertz cavity can be coupled to the spin-fluctuations in a Mott insulating antiferromagnet, which is meant to model the parent compound for a high-$T_c$ cuprate superconductor.
In our work we examined how an infrared-active phonon mode (in this case associated to the in-phase motion of the planar oxygens) can mediate the coupling between the spin-waves and the cavity.
The first scheme, outlined in Sec.~\ref{sec:polaritons}, takes spin-orbit coupling into account~\cite{Falko.2006,Lee.2020f7,Chaudhary.20200jr,Malz.2019,Cao.2015,Juraschek.2017,Talbayev.2011,Katsura.2007,Shuvaev.2010}, and is in principle present in both monolayer and bilayer cuprate systems. 
In a second scheme, we considered a coupling to the scalar exchange interaction due to linear changes of the ligand bond angle; this can only happen in a bilayer systems unless inversion symmetry is broken~\cite{Normand.1996,Sakai.1997}. 
We then proceeded to show how this phonon-mediated spin-orbit mechanism could induce hybridization of the cavity photons with the Neel magnons\textemdash forming Neel magnon-polaritons. 

Though always present, these Neel-magnon-polaritons only hybridize at finite momenta and therefore require near-field Terahertz engineering in order to be detected. 
While it presents an engineering challenge, the technologies associated to near-field coupling of Terahertz resonators has seen rapid development and is now quite promising~\cite{Zhang.2016nsw,Benea-Chelmus.2020,Yen.2004,Scalari.2012,Maissen.2014ta,Geiser.2012,Maissen.2019,Bitzer.2009,Zhang-thesis.2019,Lu.2020,Zhang.2018jbq,Fei.2011,McLeod.2014,Cvitkovic.2007,Anderson.2005,Isakov.2017,Benz.2016,Chen.2014,Chen.2003,Stinson.2014,Sternbach.2021,Sun.20205op,Jiang.2016,Yang.2016}.

We then studied the coupling to the bimagnons through phonon modulated superexchange, which is only present in the bilayer system. 
In this case, it was found that the cavity only couples to the magnon bi-linear operator in the absence of spin-orbit interactions.
We nevertheless show that this bimagnon coupling still leads to a significant dielectric response, at a frequency governed by the sum of the two relevant magnon band gaps. 
In all likelihood, when present, this response will dominate over the spin-orbit mediated mechanism and allow for the strong coupling of cavity photons to the bimagnon resonance.
Furthermore, going beyond linear response it is evident that the bimagnon coupling outlined can also be used as a parametric drive, which presents very interesting possibilities for future studies~\cite{Michael.2020,Juraschek.2020,Rajasekaran.2016,Dolgirev.2021,Bukov.2015,Malz.2019}. 

Coupling of cavity photons to antiferromagnetic spin-fluctuations, as we have proposed in this work, opens the door to the study of a number of very interesting possibilities.
One such avenue is to study whether strong coupling to a cavity can be used to enhance (or destroy) existing antiferromagnetic order, or even produce novel magnetically ordered phases~\cite{Claassen.2017,Bostrom.2020,Chiocchetta.2020}.
This would require extending our work to include effects such as strong frustration or spin-orbit coupling, both of which may lead to novel magnetic interactions.

Another extension, which is even more directly related to our work, is to study the effect of finite carrier (e.g. hole) doping~\cite{Terashige.2019}.
At finite doping charge carriers are known to strongly interact with the background antiferromagnetic order, as well as the phonon modes themselves~\cite{Normand.1996,Baldini.2020}, and may exhibit a number of competing tendencies~\cite{Sachdev.2010,Sakai.2012,Keimer.2015yde} including antiferromagnetic order, superconductivity, charge order~\cite{Peli.2017,Berg.2009vx}, and possibly even topological order~\cite{Lee.2003,Lee.2006,Anderson.1993,Sachdev.2016}. 
The possibility of manipulating this complex interplay using strong coupling to cavity photons may open a new avenue towards control of strongly correlated electronic materials~\cite{Thomas.2019}.

As a preliminary study, determining the dynamics of a single hole moving in the background of the antiferromagnetic order, in the presence of the strong cavity dressing, already presents a compelling, but daunting, theoretical effort~\cite{Normand.1996,Kane.1989,Mishchenko.2007,Lee.2006,Lee.2003,Ji.2021,Grusdt.2018,Dagotto.1994,Imada.1998}.
It would also be interesting to approach the problem from the ``Fermi-liquid" perspective, treating the antiferromagnetic order as a collective mode of the Fermi liquid~\cite{Schrieffer.1989,Abanov.2003,Carbotte.1999}.
In this case, the coupling between the plasma oscillations of the electron Fermi liquid, cavity, and antiferromagnetic collective modes are all taken to be important.
This may be important in the context of overdoped cuprates~\cite{Abanov.2003,Sachdev.2016,Sachdev.2010,Keimer.2015yde}, nickelates~\cite{Kang.2021,Krishna.2020,Jiang.2020,Sakakibara.2019,Li.201907e}, correlated oxide heterostructures~\cite{Lee.2020f7,Juraschek.2021,Schrodi.2020}, and other compounds with spin-density wave tendencies~\cite{Scalapino.1986,Sachdev.2016,Paglione2010}. 

More broadly, the mechanism we discuss can easily be generalized to other kinds of magnetic systems.
Perhaps some of the more interesting candidates would be various realizations of spin-liquids in strongly-correlated materials~\cite{Potter.2013,Claassen.2017,Bulaevskii.2008,Chiocchetta.2020,Savary.2016,Pilon.2013,Czajka2021}, iridate compounds with large spin-orbit interactions~\cite{Seifert.2019,Kim1329,KimPRL2008,Bertinshaw2019}, electron-doped cuprates~\cite{Sarkar.2020,Greene2010,Greene2020,nrp2021,sarkar2021}, or two-dimensional van der Waals materials~\cite{Mandal.2020,MacNeill.2019,Cao.2018w2g,Basov.2016,Sunku.2018,Sternbach.2021,Das.2013}.
Generically, we find that the cavity coupling is qualitatively enhanced by large spin-orbit interactions, and also in the presence of bilayer unit cells, which naturally lead to the above candidates. 

Finally, we also comment that our work is also relevant within the context of ``cavity sensing" techniques~\cite{Head-Marsden.2020}, and may be useful for detecting novel properties of quantum matter~\cite{Allocca.2019,Chatterjee.2021,Poniatowski.2021,Rodriguez-Nieva.2018,Chatterjee.2019}.
This is particularly true once the system is doped away from the insulating phase, and especially within the enigmatic pseudogap region, where strong coupling to an electromagnetic resonator may afford further insight into puzzling reports of time-reversal symmetry breaking~\cite{Poniatowski.2021,Zeng.2021,Grinenko.2021} and inversion symmetry breaking~\cite{Zhao.2017,Viskadourakis.2015,Mukherjee.2012}.

\begin{acknowledgements}
The authors thank Dominik Juraschek, Fabian Menges, Ilya Esterlis, Stephan Jesse, J{\'e}r{\^o}me Faist, Andrea Cavalleri, Dmitri N. Basov, Ata{\c c} {\. I}mamo{\u g}lu, Richard Averitt, and Amir Yacoby for fruitful discussions regarding the manuscript.
Work by J.B.C., N.R.P., and P.N. was partially supported by the Quantum Science Center (QSC), a National Quantum Information Science Research Center of the U.S. Department of Energy (DOE).
J.B.C. is an HQI Prize Postdoctoral Fellow and gratefully acknowledges support from the Harvard Quantum Initiative.
N.R.P. is supported by the Army Research Office through an NDSEG fellowship. 
V.G. and A. G. were supported by NSF DMR-2037158, US-ARO Contract No.W911NF1310172, and Simons Foundation. 
P.N. is a Moore Inventor Fellow and gratefully acknowledges support through Grant GBMF8048 from the Gordon and Betty Moore Foundation.
E.D. was supported by Harvard-MIT CUA, AFOSR-MURI: Photonic Quantum Matter award FA95501610323, the ARO grant “Control of Many-Body States Using Strong Coherent Light-Matter Coupling in Terahertz Cavities,” and the Harvard Quantum Initiative.

\end{acknowledgements}

\bibliography{references}

\appendix

\section{\label{app:low-energy-hubbard}Low-Energy Theory of Hubbard Model}

Here we derive the low-energy superexchange interactions by integrating out charge fluctuations in the half-filled Hubbard model.
We use a single-band Hubbard model with both regular and spin-orbit hopping, of the form 
\begin{equation}
    H =  -\sum_{j} \sum_{\bm \delta} c_{j+\bm \delta \alpha}^\dagger t^{\alpha\beta}_{\bm \delta} c_{j\beta }+ U \sum_{j}n_{j\uparrow}n_{j\downarrow}
\end{equation}
where $\bm \delta$ label the nearest-neighbor lattice sites of site $j$ and $\alpha,\beta$ are the spin-quantum numbers of the electrons.
We assume that the Hubbard $U \gg t$ so that we are well within the localized regime.
The hopping matrix elements $t_{\bm \delta}^{\alpha\beta}$ represent the amplitude for an electron of spin $\beta$ to hop to the neighboring site in the $\bm \delta$ direction, with a final spin state of $\beta$.
Time-reversal symmetry implies that we can write these as 
\begin{equation}
    t_{\bm \delta}^{\alpha\beta} = t_{\bm \delta} \delta_{\alpha\beta} + i \bm\lambda_{\bm \delta}\cdot \bm\sigma_{\alpha\beta}
\end{equation}
with $t,\bm\lambda$ real parameters. 
Hermiticity and translational symmetry constrain these to obey $t_{\bm \delta} = t_{-\bm \delta}$ and $\bm\lambda_{\bm\delta} = -\bm\lambda_{-\bm \delta}$. 

We now implement the Schrieffer-Wolf elimination of the doublon states.
Let $\mathcal{P}_n$ be the projector onto the subspace with $n$ total doublons.
We will restrict to the limit where only $\mathcal{P}_0, \mathcal{P}_1$ are needed, which is sufficient for second-order perturbation theory.
We can partition the Hamiltonian into an effective block structure as 
\begin{equation}
    H = \begin{pmatrix}
    0 & \mathcal{P}_0 H \mathcal{P}_1 \\
    \mathcal{P}_1 H \mathcal{P}_0 & U  \\ 
    \end{pmatrix},
\end{equation}
where we have used the fact that at exactly half-filling $\mathcal{P}_0 H \mathcal{P}_0 = 0$ and $\mathcal{P}_1 H \mathcal{P}_1 =  U \mathcal{P}_1$ to lowest order in $t/U$.
We then can adiabatically eliminate the doublon subspace, valid at low-frequencies $\omega \ll U$, such that we end up with an effective Hamiltonian (which acts on the half-filled manifold)
\begin{equation}
    H_{\rm eff} = -\frac{1}{U}  \sum_{j} \sum_{\bm \delta} c_{j+\bm \delta \alpha}^\dagger t^{\alpha\beta}_{\bm \delta} c_{j\beta } \mathcal{P}_1 \sum_{j} \sum_{\bm \delta} c_{j+\bm \delta \alpha}^\dagger t^{\alpha\beta}_{\bm \delta} c_{j\beta }.
\end{equation}
We evaluate this by choosing a site $j$ and summing over all virtual processes which go from $j$ to $j+\bm \delta$ and back. 
Summing over all of these terms exactly once will reproduce the superexchange model without double counting. 
This produces 
\begin{equation}
    H_{\rm eff} = -\frac{1}{U} \sum_{j\bm\delta}  c_{j}^\dagger \left[ t_{ -\bm \delta} + i\bm\lambda_{-\bm \delta} \cdot\bm\sigma\right] c_{j +\bm \delta } c_{j+\bm \delta}^\dagger\left[  t_{\bm \delta} + i\bm\lambda_{\bm \delta} \cdot\bm\sigma\right] c_{j} 
\end{equation}
where the spin indices are suppressed, but understood to act according to usual matrix-multiplication.
The Hamiltonian is understood as being evaluated on the half-filling space, in which the projectors become redundant and are dropped.
We evaluate this by first focusing on a single bond, where we have the term 
\begin{equation}
H_{12} =    -\frac{1}{U} c_1^\dagger \left( t-i\bm \lambda \cdot\bm\sigma\right)c_2 c_2^\dagger \left( t + i\bm\lambda \cdot\bm \sigma\right)c_1 .
\end{equation}
We can evaluate this by rotating to the $\bm \lambda$ axis, such that we have 
\begin{equation}
H_{12} = -\frac{1}{U} c_1^\dagger \left( t-i\lambda\sigma_3\right)c_2 c_2^\dagger \left( t + i\lambda\sigma_3\right)c_1 .
\end{equation}
There are now three terms; the regular super-exchange term 
\begin{equation}
H_{12}^0  = -\frac{t^2}{U} c_1^\dagger c_2 c_2^\dagger c_1 ,
\end{equation}
the antisymmetric Dzyaloshinskii-Moriya exchange 
\begin{equation}
H_{12}^1  = -i\frac{t\lambda }{U}\left(  c_1^\dagger c_2 c_2^\dagger\sigma_3  c_1  - c_1^\dagger \sigma_3 c_2c_2^\dagger c_1 \right),
\end{equation}
and the symmetric anisotropy 
\begin{equation}
H_{12}^2  = -\frac{\lambda^2}{U} c_1^\dagger \sigma_3  c_2 c_2^\dagger\sigma_3  c_1 . 
\end{equation}
Up to constants which are trivial in the half-filling sector, these can be evaluated to be 
\begin{subequations}
\begin{align}
& H_{12}^0  = 2\frac{t^2}{U}\hat{\mathbf{S}}_1\cdot \hat{\mathbf{S}}_2  \\ 
& H_{12}^1  = -2i\frac{t\lambda}{U}\left[ \hat{S}^+_1\hat{S}_2^-  - \hat{S}_1^- \hat{S}_2^+\right] \\ 
& H_{12}^2  = 2\frac{\lambda^2}{U}\left( \hat{S}_1^3 \hat{S}_2^3 - \frac12 \hat{S}_1^+ \hat{S}_2^- - \frac12 \hat{S}_1^- \hat{S}_2^+\right). \\ 
\end{align}
\end{subequations}
These are expressed in a covariant manner in terms of the spin-orbit vector as 
\begin{subequations}
\begin{align}
& H_{12}^0  = 2\frac{t^2}{U}\hat{\mathbf{S}}_1\cdot \hat{\mathbf{S}}_2  \\ 
& H_{12}^1  = -4\frac{t}{U}\bm{\lambda} \cdot \left( \hat{\mathbf{S}}_1\times \hat{\mathbf{S}}_2\right) \\ 
& H_{12}^2  = 2\frac{1}{U}\left[  2\left(\bm{\lambda}\cdot\hat{\mathbf{S}}_1\right)\left(\bm{\lambda}\cdot\hat{\mathbf{S}}_2 \right)  - \lambda^2 \hat{\mathbf{S}}_1\cdot \hat{\mathbf{S}}_2 \right] . 
\end{align}
\end{subequations}
This then brings us to the full superexchange Hamiltonian 
\begin{multline}
    H = \sum_{j\bm \delta} \frac{2(t^2 - \bm\lambda^2_{\bm \delta})}{U}\hat{\mathbf{S}}_j \cdot\hat{\mathbf{S}}_{j+\bm \delta} \\
    -\frac{4t}{U}\bm\lambda_{\bm \delta} \cdot \left(\hat{\mathbf{S}}_j\times \hat{\mathbf{S}}_{j+\bm \delta}\right) \\
    + \frac{4}{U} \left( \bm\lambda_{\bm \delta}\cdot \hat{\mathbf{S}}_j\right)\left( \bm\lambda_{\bm \delta}\cdot\hat{\mathbf{S}}_{j+\bm \delta}\right).
\end{multline}
Using the Rashba spin-orbit texture amounts to substituting $\bm \lambda_{\bm \delta} = (-1)^{\ell}\lambda \bm \delta \times \mathbf{e}_z$, with $\ell = u,d$ indexing the layer of the YBCO bilayer unit cell, as in the main text.
In fact, each bond is counted twice in this expression such that if we perform the sum over bonds we must double the exchange interactions.

\subsection{Phonon Coupling}
We now derive the coupling of the phonon displacement to the spin system by downfolding the electron-phonon Hamiltonian in the same manner. 
We include the spin-orbit and magnetostrictive couplings by taking  
\begin{equation}
    t \to t + \alpha Q ,\quad \lambda \to \lambda + \kappa Q 
\end{equation}
in the above formulae and expanding to linear order in $Q$.
To leading order, this leads to a spin-phonon interaction for each pair of sites $,1,2$ of 
\begin{multline}
    H_{\rm int} = \frac{8 t \alpha}{U} \hat{Q}_{12} \mathbf{\hat{S}}_1 \cdot\mathbf{\hat{S}}_2 - \frac{8\alpha}{U}\hat{Q}_{12}\mathbf{\lambda}\cdot \mathbf{\hat{S}}_1\times\mathbf{\hat{S}}_2  \\
    - \frac{8t\kappa}{U \lambda} \hat{Q}_{12} {\bm\lambda} \cdot \mathbf{\hat{S}}_1\times\mathbf{\hat{S}}_2  - \frac{8 \lambda \kappa  }{U}\hat{Q}_{12} \mathbf{S}_1\cdot\mathbf{S}_2 + \frac{16 \kappa }{U \lambda} \hat{Q}_{12} {\bm \lambda}\cdot\mathbf{\hat{S}}_1 {\bm \lambda}\cdot\mathbf{\hat{S}}_2 .
\end{multline}
The first two terms come from linear variation of the scalar hopping $t$, while the last three come from the linear variation in the spin-orbit coupling. 
We anticipate the hierarchy of coupling constants $8t\alpha/U \gg 8 \lambda \alpha/ U \gtrsim 8t\kappa /U \gg 8 \lambda \kappa /U$.

\section{\label{app:EP}Easy-Plane Model}
Here we show how to diagonalize the easy-plane spin-wave Hamiltonian.
We assume a classical N\'eel order which lies in the $x-y$ plane at an angle $\theta$ from the $x$ axis, and further choose the $x$ and $y$ axes to align with the $a$ and $b$ crystallographic axes, respectively. 
We then introduce the ``1,2,3" coordinate system, such that 
\begin{equation}
    \mathbf{e}_3 = \cos \theta \mathbf{e}_x + \sin \theta \mathbf{e}_y 
\end{equation}
defines the axis along which the N\'eel order lies. 
Then we choose the transverse $1$ and $2$ directions such that 
\begin{equation}
    \mathbf{e}_1 \times \mathbf{e}_2 = \mathbf{e}_3,
\end{equation}
i.e. the coordinate system is right-handed. 
One such choice is 
\begin{subequations}
\begin{align}
    \mathbf{e}_1 &= \mathbf{e}_z \\
    \mathbf{e}_2 &= \sin \theta \mathbf{e}_x-\cos\theta \mathbf{e}_y \\
    \mathbf{e}_3 &= \cos \theta \mathbf{e}_x + \sin \theta \mathbf{e}_y .
\end{align}
\end{subequations}
In this basis, we write the Holstein-Primakoff expansion as 
\begin{equation}
    \hat{S}^3_{j} = \begin{cases}
    \frac12 - a_{j}^\dagger a_{j} & j \in A \\
    b_{j}^\dagger b_{j} - \frac12 & j \in B. \\
    \end{cases}
\end{equation}
On the two sublattices we then have a linear spin-wave expansion of 
\begin{equation}
    \hat{S}^+_{j} = \hat{S}^1_{j} + i \hat{S}^2_{j} = \begin{cases}
    a_{j} & j \in A \\
    b_{j}^\dagger  & j \in B. \\
    \end{cases}
\end{equation}

The Hamiltonian can be then written in terms of the Holstein-Primakoff bosons as 
\begin{multline}
    H_0 = \sum_{\bf q} \left(a_{\bf q }^\dagger , b_{\bf -q} \right)\begin{pmatrix}
    2J  & (2J-\Gamma ) \gamma_{\bf q}^s \\ 
    (2J-\Gamma )\gamma_{\bf q}^s & 2J  \\ 
    \end{pmatrix}\begin{pmatrix}
    a_{\bf q} \\
    b_{-\bf q}^\dagger \\
    \end{pmatrix} \\
    -\sum_{\bf q} \Gamma \gamma_{\bf q}^s \left( a_{\bf q}^\dagger b_{\bf q} + a_{\bf q} b_{\bf q}^\dagger\right).
\end{multline}

Next, we introduce the four component Nambu spinor 
\begin{equation}
    \Psi_{\bf q} = \begin{pmatrix}
    a_{\bf q}\\
    b_{\bf q}\\ 
    a_{-\bf q}^\dagger \\
    b_{-\bf q}^\dagger 
    \end{pmatrix},
\end{equation}
and the associated Nambu Pauli matrices $\tau_a$ (which act on the $(X,X^\dagger)$ doublet), as well as the sublattice Pauli matrices $\zeta_a$ (which act on the $(a,b)$ doublet).
We then have 
\begin{equation}
    H = \frac12 \sum_{\bf q}\Psi_{\bf q}^\dagger \mathbb{M}_{\bf q} \Psi_{\bf q}
\end{equation}
with the quadratic form 
\begin{equation}
    \mathbb{M}_{\bf q} = (2J - \Gamma \gamma_{\bf q }^s \zeta_1 ) + \tau_1 (2J - \Gamma \gamma_{\bf q }^s \zeta_1 ).
\end{equation}
The central object of interest is the magnon Green's function, which is obtained by diagonalizing the kernel 
\begin{equation}
    \mathbb{D}^{-1}(q)  = i\omega_m \tau_3 - \mathbb{M}_{\bf q}.
\end{equation}
This can be diagonalized by first going to the eigenbasis of $\zeta_1$, which diagonalizes the sublattice eigenvalue $\zeta = \pm 1$.
We then diagonalize the Nambu matrices in order to obtain the magnon dispersion relation,
\begin{equation}
    \Omega_{\bf p\zeta } = \sqrt{ ( 2J - \zeta\Gamma\gamma_{\bf p}^s)^2 - (2J - \Gamma)^2(\gamma_{\bf p}^s)^2 }.
\end{equation}
While the two magnon bands are degenerate for the Heisenberg model, they now become split in the presence of anisotropy. 
At long wavelengths, one is gapped with gap 
\begin{equation}
    \Omega_{\bf p,-} \sim \sqrt{ 8J\Gamma} + ...,
\end{equation}
while the other one remains gapless, with a spin-wave velocity of 
\begin{equation}
    \Omega_{\bf p,+} \sim \sqrt{2J(2J-\Gamma)}|\bf p|.
\end{equation}

We now determine the coupling to the electric field through the DM-type interaction for the case of N\'eel order in the $x-y$ plane. 
We consider the coupling 
\begin{equation}
    H_{\rm int} = \sum_{j\in A, \bm \delta} g \hat{E}^z_{j,j+\bm \delta} \left( \mathbf{e}_z\times \bm{\delta} \right) \cdot \left(\mathbf{\hat{S}}_{j}\times \hat{\mathbf{S}}_{j+\bm \delta} \right).
\end{equation}
We first express this in momentum space and separate out the linear and quadratic contributions as  
\begin{equation}
    H_{\rm int}^{(2)} = \sum_{\bf q} g \hat{E}^z_{\bf q} \sum_{\bm \delta} \left[ \left( \mathbf{e}_z \times \bm \delta \right) \cdot \mathbf{e}_2 \right] e^{i\frac{\mathbf{q}\cdot \bm \delta}{2}} \left( \mathbf{\hat{S}}_{j} \times \mathbf{\hat{S}}_{j+\bm \delta} \right)_{2,\mathbf{q}},
\end{equation}
\begin{equation}
    H_{\rm int}^{(3)} = \sum_{\bf q} g \hat{E}^z_{\bf q} \sum_{\bm \delta} \left[ \left( \mathbf{e}_z \times \bm \delta \right) \cdot \mathbf{e}_3 \right] e^{i\frac{\mathbf{q}\cdot \bm \delta}{2}} \left( \mathbf{\hat{S}}_{j} \times \mathbf{\hat{S}}_{j+\bm \delta} \right)_{3,\mathbf{q}},
\end{equation}
With 
\begin{equation}
    \left(\mathbf{\hat{S}}_j \times \mathbf{\hat{S}}_{j+\bm \delta} \right)_{\bf q} = \sum_{j}\frac{e^{i\mathbf{q}\cdot\mathbf{R}_j }}{\sqrt{\rm Vol.}} \left( \mathbf{\hat{S}}_j \times \hat{\mathbf{S}}_{j+\bm \delta} \right).
\end{equation}
We find the resulting interaction of $H_{\rm int} = H_{\rm int}^{(2)} + H_{\rm int}^{(3)} $ with 
\begin{subequations}
\begin{align}
& H_{\rm int}^{(2)} =\sum_{\bf p} \frac{ig}{2} \hat{E}^z_{-\bf p} \left(\cos \theta \sin \frac{p_x}{2} + \sin \theta \sin \frac{p_y}{2} \right) \left( a_{\bf q} - b_{\bf q} \right)+\textrm{h.c.}\\
& H_{\rm int}^{(3)} = \sum_{\bf q,p}\frac{g\hat{E}^z_{-\bf q} }{\sqrt{\rm Vol.}}\left( \sin \theta \sin p_x - \cos \theta \sin p_y \right) a_{\bf p -\frac{q}{2}}^\dagger b_{\bf -p - \frac{q}{2}}^\dagger + \textrm{h.c.} 
\end{align}
\end{subequations}
and in particular, if we project onto $\mathbf{q} =0$ for the electric field, we find the first term vanishes linearly with $\bf q$, while the second term recovers a finite homogeneous value.

\section{\label{app:bilayer}Bilayer Hamiltonian}

Here we calculate the spin-wave Hamiltonian for the bilayer system, assuming N\'eel order lies in the $a-b$ plane.
We use the same ``1,2,3" coordinates which were introduced in Appendix~\ref{app:EP} for the case of easy-plane N\'eel order.
The only further complication is due to the bilayer structure.

We focus on the case of the easy-plane N\'eel order since, following Ref.~\cite{Bonesteel.1993}, we expect the in-plane N\'eel order to be stabilized by the frustrated anisotropies.
Heuristically, this can be understood from the fact that, while different bond directions in the $a-b$ plane frustrate the anisotropies of each other, when the spin lies in the $x-y$ plane, it gains {\it some} anisotropy energy.
On the other hand, when it points along $z$ it gains no anisotropy energy.
Therefore, the in-plane order is stabilized. 
We would further expect that, because spin-rotation symmetry is broken explicitly by the frustrated spin-orbit coupling, a similar order-by-disorder mechanism ultimately opens a spin-wave gap even though classically the frustration produces a rotationally invariant interaction.

The only difference from the single-layer case is that we must also alternate the sublattice between the two bilayers, so that we write the Holstein-Primakoff expansion as 
\begin{equation}
    \hat{S}^3_{j\ell} = \begin{cases}
    \frac12 - a_{j\ell}^\dagger a_{j\ell} & (j,\ell) \in A \\
    b_{j\ell}^\dagger b_{j\ell} - \frac12 & (j,\ell) \in B. \\
    \end{cases}
\end{equation}
We use the notation $(j,\ell) \in A,B$ to indicate that the sites on the $A$ sublattice in the upper layer are in the $B$ sublattice in the lower layer, and vice-versa. 
On the two sublattices we have linear spin-wave expansions of 
\begin{equation}
    \hat{S}^+_{j\ell} = \hat{S}^1_{j\ell} + i \hat{S}^2_{j\ell} = \begin{cases}
    a_{j\ell} & (j,\ell) \in A \\
    b_{j\ell}^\dagger  & (j,\ell) \in B. \\
    \end{cases}
\end{equation}
We now compute all the needed operators in terms of the local axes,  expand in terms of the Holstein-Primakoff bosons, and express them in momentum space.

\subsection{Interlayer Exchange}
We first express the interlayer coupling in terms of the local axes.
This is easy because it is a spin-scalar and therefore is basis independent. 
As such, we have 
\begin{multline}
    \mathbf{\hat{S}}_{ju}\cdot\mathbf{\hat{S}}_{jd} = \hat{S}_{ju}^3 \hat{S}_{jd}^3 +\frac12 \hat{S}_{ju}^+ \hat{S}_{jd}^- + \frac12 \hat{S}_{ju}^- \hat{S}_{jd}^+ \\
    = \begin{cases} 
    \frac12 \left[ a_{ju}^\dagger a_{ju} + b_{jd}^\dagger b_{jd} + a_{ju} b_{jd} + a_{ju}^\dagger b_{jd}^\dagger \right] & j\in A \\
    \frac12 \left[ b_{ju}^\dagger b_{ju} + a_{jd}^\dagger a_{jd} + b_{ju} a_{jd} + b_{ju}^\dagger a_{jd}^\dagger \right] & j\in B, \\
    \end{cases}
\end{multline}
where the two cases correspond to the case where $j$ is such that $\mathbf{R}_j$ is in the A sublattice on the upper layer (and therefore on the $B$ sublattice on the lower layer). 
This is the only term which couples the two layers. 
The remaining terms are now evaluated for the upper layer.
To obtain the terms on the lower layer we take $\lambda \to -\lambda$ and switch the two sublattices with each other. 

To translate this to momentum space, we note that there is no dispersion from this term, only an ``interband" coupling.
We then obtain 
\begin{multline}
H_{ud}=J_{ud}\sum_{j\in A,B} \mathbf{\hat{S}}_{ju}\cdot\mathbf{\hat{S}}_{jd} = \\
\frac12 J_{ud}\sum_{\bf p \ell} \left( a_{\bf p \ell}^\dagger a_{\bf p \ell} + b_{\bf p \ell }^\dagger b_{\bf p\ell} \right)\\
+\frac12 J_{ud}\sum_{\bf p} \left(a_{\mathbf{p} u}^\dagger ,b_{-\mathbf{p} u}\right) \begin{pmatrix}
0 & 1\\
1 & 0\\
\end{pmatrix}\begin{pmatrix}
a_{\mathbf{p} d} \\
b_{-\mathbf{p} d }^\dagger \\
\end{pmatrix} \\
+\frac12 J_{ud}\sum_{\bf p}\left(a_{\mathbf{p} d}^\dagger ,b_{-\mathbf{p} d}\right) \begin{pmatrix}
0 & 1\\
1 & 0\\
\end{pmatrix}\begin{pmatrix}
a_{\mathbf{p} u} \\
b_{-\mathbf{p} u }^\dagger \\
\end{pmatrix}.\\
\end{multline}

\subsection{Scalar Exchange}
We now evaluate the scalar exchange interaction for each layer.
Again, this is simplified since it is invariant upon changing to the local axes. 
The sum over all neighboring bonds is implemented by summing over $j \in A$ and all the $\bm \delta$ vectors.
We then have 
\begin{equation}
\mathbf{\hat{S}}_{j\ell} \cdot\mathbf{\hat{S}}_{j+\bm \delta \ell} = \frac12 \left[ a_{j\ell}^\dagger a_{j\ell} + b_{j+\bm \delta \ell}^\dagger b_{j+\bm \delta \ell} + a_{j\ell}b_{j+\bm \delta \ell}+ a_{j\ell}^\dagger b_{j+\bm \delta \ell}^\dagger \right]   
\end{equation}
for the scalar term, which is valid for both layers provided we are careful to only sum over sites in the A sublattice, which are translated relative to each other between each respective layer. 
We translate this into momentum space to find the Hamiltonian 
\begin{equation}
H_0 = 2J_0 \sum_{\ell}\sum_{\bf p} \left[ a_{\bf p\ell}^\dagger a_{\bf p\ell} + b_{\bf p \ell}^\dagger b_{\bf p \ell} + \gamma_{\bf p} a_{\bf p\ell}b_{-\bf p  \ell}+\gamma_{\bf p} a_{\bf p\ell}^\dagger b_{-\bf p\ell}^\dagger \right]   
\end{equation}
with usual form factor $\gamma_{\bf p} = \frac12 \left( \cos p_x + \cos p_y \right)$ and $J_0 = 4 (t^2 - \lambda^2)/U$ the scalar exchange.

\subsection{Vector Exchange}
The vector exchange is obtained by expanding in the $1,2,3$ basis as 
\begin{widetext}
\begin{equation}
    \mathbf{\hat{S}}_j \times \mathbf{\hat{S}}_{j+\bm \delta} = \mathbf{e}_3 \frac{\hat{S}^-_j \hat{S}^+_{j+\bm \delta} - \hat{S}^+_j \hat{S}^-_{j+\bm \delta} }{2i} + \hat{S}_j^3 \left( i \mathbf{e}_+ \hat{S}^+_{j+\bm \delta} -  i \mathbf{e}_- \hat{S}^-_{j+\bm \delta} \right) - \left( i \mathbf{e}_+ \hat{S}^+_{j} -  i \mathbf{e}_- \hat{S}^-_{j} \right) \hat{S}_{j+\bm \delta}^3 ,
\end{equation}
with $\mathbf{e}_\pm = \frac12 (\mathbf{e}_1 \mp i \mathbf{e}_2)$.
We expand this operator up to quadratic order in the Holstein-Primakoff bosons.
We obtain 
\begin{equation}
    \mathbf{\hat{S}}_j \times \mathbf{\hat{S}}_{j+\bm \delta} = \mathbf{e}_3 \frac{a^\dagger_j b^\dagger_{j+\bm \delta} - a_j b_{j+\bm \delta} }{2i} + \frac12 \left( i \mathbf{e}_+ b^\dagger_{j+\bm \delta} -  i \mathbf{e}_- b_{j+\bm \delta} \right) +\frac12 \left( i \mathbf{e}_+ a_{j} -  i \mathbf{e}_- a^\dagger_{j} \right). 
\end{equation}

In particular, we need the projection of this operator onto the vector $\mathbf{e}_z \times \bm \delta $. 
We obtain 
\begin{equation}
\mathbf{\hat{S}}_{j}\times \mathbf{\hat{S}}_{k} \cdot (\mathbf{e}_z \times \bm \delta ) \\
=\begin{cases}
\sin\theta \left(\frac{a^\dagger_j b^\dagger_{j+\bm \delta} - a_j b_{j+\bm \delta} }{2i} \right)  -\frac{\cos\theta}{4} \left(b_{j+\bm \delta}^\dagger + b_{j+\bm \delta} + a_j + a_j^\dagger\right)  & \bm \delta = \mathbf{e}_x \\
-\cos \theta \left(\frac{a^\dagger_j b^\dagger_{j+\bm \delta} - a_j b_{j+\bm \delta} }{2i} \right) - \frac{\sin \theta}{4} \left(b_{j+\bm \delta}^\dagger + b_{j+\bm \delta} + a_j + a_j^\dagger\right)  & \bm \delta = \mathbf{e}_y .\\
\end{cases}
\end{equation}
\end{widetext}
We have suppressed the layer index for brevity (since at the moment the result is the same for both). 
To express this in momentum space, we note that the linear term will cancel since the sum involves terms with both $\bm{\delta}$ and $-\bm \delta$. This leaves 
\begin{multline}
    H_{1} = -\sum_{j\in A,\ell} (-1)^{\ell}\frac{8t\lambda}{U} \Bigg[ \\
    \sin \theta \left( \frac{ a_{j\ell}^\dagger b_{j+x\ell}^\dagger-a_{j\ell}b_{j+x\ell} }{2i} - \frac{ a_{j\ell}^\dagger b_{j-x\ell}^\dagger-a_{j\ell}b_{j-x\ell} }{2i}\right) \\
    - \cos \theta \left( \frac{ a_{j\ell}^\dagger b_{j+y\ell}^\dagger-a_{j\ell}b_{j+y\ell} }{2i} - \frac{ a_{j\ell}^\dagger b_{j-y\ell}^\dagger-a_{j\ell}b_{j-y\ell} }{2i}\right) \Bigg].\\
\end{multline}
Going to momentum space we find 
\begin{multline}
    H_{1} = \sum_{\ell \mathbf p} -(-1)^{\ell}\frac{8t\lambda}{U} \left[ \sin \theta \sin p_x - \cos \theta  \sin p_y \right] \\
    \times\left( a_{\mathbf{p} \ell}b_{-\mathbf{p}\ell} + b_{-\mathbf{p}\ell}^\dagger a_{\mathbf{p}\ell}^\dagger \right).\\
\end{multline}

\subsection{Tensor Exchange}
The tensor exchange is 
\begin{multline}
\mathbf{\hat{S}}_{j}\cdot (\mathbf{e}_z \times \bm \delta) (\mathbf{e}_z \times \bm \delta )\cdot \mathbf{\hat{S}}_{k}  \\
= \begin{cases}
    \left(-\cos \theta \hat{S}_{j}^2 + \sin \theta \hat{S}_{j}^3  \right)\left( -\cos \theta \hat{S}_{k}^2 + \sin \theta \hat{S}_k^3 \right) & \bm \delta = \mathbf{e}_x \\
    \left(-\sin \theta \hat{S}_{j}^2 - \cos \theta \hat{S}_{j}^3  \right)\left( -\sin \theta \hat{S}_{k}^2 - \cos \theta \hat{S}_k^3 \right) & \bm \delta = \mathbf{e}_y .\\
     \end{cases}
\end{multline}
Next, we express this in terms of the angular momentum ladder operators $S^\pm = S^1 \pm i S^2 $. 
We have 
\begin{widetext}
\begin{equation}
\mathbf{\hat{S}}_{j\ell}\cdot (\mathbf{e}_z \times \bm \delta) (\mathbf{e}_z \times \bm \delta )\cdot \mathbf{\hat{S}}_{j+\bm \delta \ell} = \begin{cases}
    \left(-\cos \theta \frac{\hat{S}_{j\ell}^+ - \hat{S}^-_{j\ell}}{2i} + \sin \theta \hat{S}_{j\ell}^3  \right)\left( -\cos \theta \frac{\hat{S}_{j+\bm \delta \ell}^+ - \hat{S}^-_{j+\bm \delta\ell}}{2i} + \sin \theta \hat{S}_{j+\bm \delta \ell}^3 \right) & \bm \delta = \mathbf{e}_x \\
    \left(-\sin \theta \frac{\hat{S}_{j\ell}^+ - \hat{S}^-_{j\ell}}{2i} - \cos \theta \hat{S}_{j\ell}^3  \right)\left( -\sin \theta \frac{\hat{S}_{j+\bm \delta \ell}^+ - \hat{S}^-_{j+\bm \delta\ell}}{2i} - \cos \theta \hat{S}_{j+\bm \delta\ell}^3 \right) & \bm \delta = \mathbf{e}_y .\\
     \end{cases}    
\end{equation}
This is now expressed in terms of the Holstein-Primakoff bosons, up to quadratic order. 
This is slightly more complicated because there are cross terms between $S^3$, which contains the zeroth and second order terms, and the transverse terms which contain linear terms.
We drop terms which are overall constants, such that 
\begin{multline}
\mathbf{\hat{S}}_{j\ell}\cdot (\mathbf{e}_z \times \bm \delta) (\mathbf{e}_z \times \bm \delta )\cdot \mathbf{\hat{S}}_{j+\bm \delta \ell} \\
= \begin{cases}
-\frac14\cos^2 \theta \left( a_{j\ell} - a_{j\ell}^\dagger \right) \left( b_{j+\bm \delta \ell}^\dagger - b_{j+\bm \delta \ell} \right) + \frac12 \sin^2\theta\left(a_{j\ell}^\dagger a_{j\ell} + b_{j+\bm \delta \ell}^\dagger b_{j+\bm \delta \ell} \right)+ \frac{1}{4i}\sin\theta \cos\theta \left( a_{j\ell} - a_{j\ell }^\dagger - b_{j+\bm \delta \ell}^\dagger + b_{j+\bm \delta \ell}\right) & \bm \delta = \mathbf{e}_x \\
-\frac14\sin^2 \theta \left( a_{j\ell} - a_{j\ell}^\dagger \right) \left( b_{j+\bm \delta \ell}^\dagger - b_{j+\bm \delta \ell} \right) + \frac12 \cos^2\theta\left(a_{j\ell}^\dagger a_{j\ell} + b_{j+\bm \delta \ell}^\dagger b_{j+\bm \delta \ell} \right)-\frac{1}{4i}\sin\theta \cos\theta \left( a_{j\ell} - a_{j\ell }^\dagger - b_{j+\bm \delta \ell}^\dagger + b_{j+\bm \delta \ell}\right) & \bm \delta = \mathbf{e}_y.\\
\end{cases}    
\end{multline}
\end{widetext}
We now pass to momentum space again.
Just like in the vector coupling, the linear term will cancel because the $x$ and $y$ anisotropies enter with opposite signs in the linear part. 
Then we observe that of the two quadratic terms, the one which conserves boson number, when summed over the four different bond values, will give $a_{j\ell}^\dagger a_{j\ell} + b_{j+\bm \delta \ell}^\dagger b_{j+\bm \delta}$, independent of the N\'eel ordering vector.
Therefore, there is one contribution which is simply
\begin{equation}
    H_{2,1} = \frac{8\lambda^2}{U} \sum_{\bf p \ell} a_{\bf p \ell}^\dagger a_{\bf p \ell} + b_{\bf p \ell}^\dagger b_{\bf p \ell}.
\end{equation}
The other contribution comes from the first term and in real space reads 
\begin{widetext}
\[
H_{2,2} = -\frac{2\lambda^2}{U} \sum_{j\in A, \ell }\left[ \cos^2 \theta \left( a_{j\ell} - a_{j\ell}^\dagger \right) \left( b_{j+x \ell}^\dagger - b_{j+x \ell} +b_{j-x \ell}^\dagger - b_{j-x \ell} \right) + \sin^2 \theta \left( a_{j\ell} - a_{j\ell}^\dagger \right) \left( b_{j+y \ell}^\dagger - b_{j+y \ell}+b_{j-y \ell}^\dagger - b_{j-y \ell} \right)\right].
\]
In momentum space, this reads 
\begin{equation}
    H_{2,2} = -\frac12 \Gamma \sum_{ \bf p\ell } (b_{\bf p \ell}^\dagger - b_{-\bf p\ell}) (a_{\bf p\ell} - a^\dagger_{-\bf p \ell}) \left( \cos p_x \cos^2 \theta + \cos p_y \sin^2 \theta \right)
\end{equation}
with $\Gamma = 8\lambda^2 /U$.
\end{widetext}

\subsection{Full Bilayer Model}
We now present the full bilayer Hamiltonian in momentum space, to second order in the Holstein-Primakoff bosons.
We have 
\begin{widetext}
\begin{multline}
H =  \frac12 J_{ud} \sum_{\bf p} a_{\mathbf{p} u}^\dagger b_{-\mathbf{p} d}^\dagger + a_{\mathbf{p} d}^\dagger b_{-\mathbf{p} u}^\dagger + b_{-\mathbf{p} d} a_{\mathbf{p} u} + b_{-\mathbf{p} u} a_{\mathbf{p} d}  + \frac12 J_{ud} \sum_{\bf p \ell} a_{\bf p \ell}^\dagger a_{\bf p\ell} +b_{\bf p\ell}^\dagger b_{\bf p\ell} \\
+\sum_{\ell \bf p}\left( a_{\bf p\ell}^\dagger, b_{-\bf p \ell}\right)\begin{pmatrix}
2J_0 +\Gamma & 2J_0\gamma^s_{\bf p}  - D_\ell \gamma^p_{\mathbf{p}}(\theta) + \frac12\Gamma \tilde{\gamma}^d_{\bf p}(\theta) \\
2J_0\gamma^s_{\bf p} - D_\ell \gamma^p_{\bf p}(\theta)  + \frac12 \Gamma \tilde{\gamma}^d_{\bf p}(\theta) & 2J_0 + \Gamma \\
\end{pmatrix}
\begin{pmatrix}
a_{\bf p\ell} \\
b_{-\bf p\ell}^\dagger\\
\end{pmatrix}  - \frac12 \sum_{\bf p\ell} \Gamma \tilde{\gamma}^d (\theta) \left( a_{\bf p\ell}^\dagger b_{\bf p\ell} + b_{\bf p\ell}^\dagger a_{\bf p\ell} \right).
\end{multline}
Here we have introduced multiple form-factors (which are dependent on the N\'eel order angle), which are defined as  
\begin{subequations}
\begin{align}
& \gamma^s_{\bf p} = \frac12 \left( \cos p_x + \cos p_y \right) \\
& \gamma^p_{\bf p}(\theta) = \sin \theta \sin p_x - \cos \theta \sin p_y \\
& \tilde{\gamma}^d_{\bf p}(\theta) =  \cos p_x \cos^2 \theta +  \cos p_y \sin^2 \theta = \gamma_{\bf p}^s + \cos 2\theta \gamma^d_{\bf p} \\
& \gamma_{\bf p}^d = \frac12 (\cos p_x - \cos p_y )
\end{align}
\end{subequations}
and the exchange constants are, in terms of the Hubbard model parameters, 
\begin{subequations}
\begin{align}
J_0 = 4 \frac{t^2 - \lambda^2}{U} \\
D_\ell = 8 \frac{t \lambda}{U} (-1)^\ell \\
\Gamma = 8 \frac{\lambda^2}{U} .
\end{align}
\end{subequations}

It will also be relevant to include the matrix elements of the Dzyaloshinskii-Moriya-type interaction with the electric field, which takes the form 
\begin{equation}
    H_{\rm int} = \sum_{j\in A\bm \delta \ell} g \hat{E}^z_{j,j+\bm \delta}\left( \mathbf{e}_z \times \bm\delta  \right) \cdot \mathbf{\hat{S}}_{j\ell}\times \mathbf{\hat{S}}_{j+\bm \delta \ell}. 
\end{equation}
This will yield the same result as previously, except we must sum over the layer index $\ell$. 
We thus find 
\begin{subequations}
\begin{align}
& H_{\rm int}^{(2)} =\sum_{\bf p \ell}\hat{E}^z_{-\bf p} i\frac{g}{2}\left(\cos \theta \sin \frac{p_x}{2} + \sin \theta \sin \frac{p_y}{2} \right) \left( a_{\bf p\ell} - b_{\bf p\ell} \right)+\textrm{h.c.}\\
& H_{\rm int}^{(3)} = \sum_{\bf q,p \ell }\frac{g\hat{E}^z_{\bf q} }{\sqrt{\rm Vol.}}\left( \sin \theta \sin p_x - \cos \theta \sin p_y \right) a_{\bf p +\frac{q}{2}\ell}^\dagger b_{\bf -p + \frac{q}{2}\ell}^\dagger + \textrm{h.c.} 
\end{align}
\end{subequations}
\end{widetext}
We can verify that inversion takes $a_{{\bf q }u}\to a_{-{\bf q} d}$ and similarly for $b$, and also takes $Q_{\bf q} \to -Q_{-\bf q}$.
Therefore, the interaction as written preserves inversion symmetry.

\subsubsection{Diagonalizing Bilayer Model}
The spin-wave Hamiltonian for the full bilayer system can be diagonalized exactly. 
Similar to the monolayer case, we introduce $\tau_a$ for Nambu space Pauli matrices and $\zeta_a$ for sublattice space Pauli matrices. 
We must now also introduce an additional layer space $l_a$ set of Pauli matrices.
In terms of the these, and the 8-component Nambu-sublattice-layer spinor $\Psi_{\bf p}$, we have the Hamiltonian 
\begin{equation}
H = \frac12 \sum_{\bf p} \Psi_{\bf p}^\dagger \hat{\mathbb{M}}_{\mathbf{p}} \Psi_{\bf p}
\end{equation}
with the matrix 
\begin{widetext}
\begin{equation}
    \hat{\mathbb{M}}_{\mathbf{p}} = \frac12 J_{ud}\left[ 1 + \tau_1 \zeta_1 l_1 \right] + 2J_0 + \Gamma - \frac12 \Gamma\left( \gamma_{\bf p}^s + \cos 2\theta \gamma_{\bf p}^d \right) \zeta_1 + \left[2J_0 \gamma_{\bf p}^s + \frac12 \Gamma \left(\gamma_{\bf p}^s + \cos 2\theta \gamma_{\bf p}^d\right)\right]\tau_1 \zeta_1 + D \gamma_{\bf p}^{p} \tau_2 \zeta_2 l_3.
\end{equation}
\end{widetext}
Our aim is to diagonalize the dynamical matrix, $\mathbb{\hat{K}}_{\bf p} = \tau_3 \mathbb{\hat{M}}_{\bf p}$, which in turn can be used to compute the magnon Matsubara Green's function by inverting the kernel 
\begin{equation}
    \hat{\mathbb{D}}_{\mathbf{p}}^{-1} = i \omega_m \tau_3 - \mathbb{\hat{M}}_{\bf p}  = \tau_3 \left( i\omega_m - \mathbb{\hat{K}}_{\bf p} \right).
\end{equation}  
We begin by noting that the sublattice parity matrix $\Pi = \zeta_1 l_1$ commutes with the Green's function and obeys $\Pi^2 = 1$.
We therefore find two two-fold degenerate representations, each of definite parity as determined by the eigenvalue of $\Pi = \pm 1$.

In order to finish diagonalizing the spin-wave kernel, we define $Z = \zeta_1$.
In each subspace of $\Pi$, we have two basis states, which can be labeled by their eigenvalue of $Z$. 
For instance, for $\Pi = +1$ the two states are $|++\rangle$ and $|--\rangle$, with the first number labeling the eigenvalue of $\zeta_1 = Z$, and the second labeling the eigenvalue of $l_1 = \Pi \zeta_1$.
The action of $\zeta_2 l_3$ on these basis states can then be mapped onto the action of $-Y$, such that the dynamical matrix assumes the form
\[
\hat{\mathbb{K}} = \tau_3 \left[ a_3 + b_3Z  \right] + i\tau_2 \left[a_2 + b_2 Z  \right] + i\tau_1 c Y  . 
\]
We know that the eigenvalues will come in particle-hole conjugate pairs, and we therefore consider squaring $\mathbb{K}$ and diagonalizing that matrix. 
We find 
\[
\hat{\mathbb{K}}^2 = (a_3 + b_3 Z)^2 - (a_2 + b_2 Z)^2 - c^2 +2i\tau_2 cb_3 X + 2\tau_3 c b_2 X .
\]
This matrix is now reduced to a Clifford algebra form, with $Z,\tau_2 X, \tau_3 X$ forming a set of three anti-commuting matrices which all square to one. 
We therefore find that the eigenvalues $\lambda$ of $\mathbb{\hat{K}} $ are
\begin{widetext}
\begin{equation}
   \lambda_{\pm} = E_{\pm}^2 =  a_3^2 + b_3 ^2 - a_2^2 - b_2^2 - c^2 \pm \sqrt{ (2a_3 b_3 - 2a_2 b_2  )^2 - (2c b_3)^2 + (2c b_2)^2 }.
\end{equation}
It follows that the energy eigenvalues are
\begin{equation}
   E_{\zeta } =  \sqrt{ a_3^2 + b_3 ^2 - a_2^2 - b_2^2 - c^2 -2\zeta \sqrt{ (a_3 b_3 - a_2 b_2  )^2 - (c b_3)^2 + (c b_2)^2 }},
\end{equation}
assuming that the energies remain real, and therefore there is no dynamical instability.
We have introduced the quantum number $\zeta = \pm$ here, which is chosen such that in the limit of $\bf p \to 0$, this coincides with the eigenvalue of $\zeta_1$, at least for a certain range of $\Gamma/J_{ud}$.
The parameters in the above formula are 
\begin{subequations}
\begin{align}
& a_3 = 2J_0 + \Gamma + \frac12 J_{ud} \\
& b_3 = -\frac12 \Gamma (\gamma_{\bf p}^s + \cos 2\theta \gamma_{\bf p}^d ) \\
& a_2 = \frac12 J_{ud} \Pi  \\
& b_2 = 2J_0 \gamma_{\bf p}^s +\frac12 \Gamma (\gamma_{\bf p}^s + \cos 2\theta \gamma_{\bf p}^d )  \\
& c = D \gamma_{\bf p}^p .
\end{align}
\end{subequations}

In particular, we have the following results as $\bf p \to 0$.  
First, as $\bf p \to 0$ we find that the quantum numbers become identified as the eigenvalue of $\zeta_1$, so we will label them as $\zeta = \pm $, even though this assignment is only accurate strictly at $\bf p = 0$.
Second, we find the energies at $\bf p = 0$ of 
\begin{equation}
E_{\Pi = \pm, \zeta = \pm}({\bf p =0}) = \sqrt{ ( 2J_0 + \Gamma + \frac12 J_{ud}  - \zeta \frac12 \Gamma)^2  - (2J_0 +\frac12 \Gamma + \zeta \Pi J_{ud}/2 )^2}.
\end{equation}
This simplifies to yield
\begin{subequations}
\begin{align}
    & E_{ ++}(0) = 0 \\
    & E_{ +-}(0) = \sqrt{2( 2J_0 + \Gamma )(\Gamma + J_{ud}) } \\    
    & E_{ -+}(0) = \sqrt{ 2(2J_0 + \Gamma/2) J_{ud} } \\
    & E_{ --}(0) = \sqrt{ 2(2J_0 + \Gamma + \frac12 J_{ud}) \Gamma  }.
\end{align}
\end{subequations}
The group velocity of the gapless pseudo-Goldstone mode is, within this approximation, given by the dispersion relation 
\begin{subequations}
\begin{align}
& E_{++}(\mathbf{p}) = \sqrt{A_{xx}p_x^2 + 2A_{xy}p_xp_y + A_{yy} p_{y}^2 } \\
& A_{xx} = \left( 2J_0 + \frac12 \Gamma + \frac12 J_{ud} \right) \left( \Gamma \cos^2 \theta + J_0 - \frac{2D^2 \sin^2 \theta}{\Gamma + J_{ud}}\right) \\
& A_{xy} = \left( 2J_0 + \frac12 \Gamma + \frac12 J_{ud} \right) \frac{2D^2}{\Gamma + J_{ud}}\sin \theta \cos \theta  \\
& A_{yy} = \left( 2J_0 + \frac12 \Gamma + \frac12 J_{ud} \right) \left( \Gamma \sin^2 \theta + J_0 - \frac{2D^2 \cos^2 \theta}{\Gamma + J_{ud}}\right).
\end{align}
\end{subequations}
This has a stability criterion that the quadratic form formed by the elements $A_{xx},A_{xy},A_{yy}$ be positive semi-definite.
For $\theta = 0$ this amounts to 
\begin{equation}
    J_{ud}/J_0 > 4 \frac{D^2 }{J_0^2} - \Gamma/J_{0}^2 \sim 6 (\lambda / t)^2 . 
\end{equation}
Therefore, for weak spin-orbit coupling (we take $\lambda /t \sim .025$), this is quite reasonable. 
\end{widetext}

\subsection{Green's Function}
We now obtain the magnon Green's function, which is used in the linear response calculations.
Writing out the Nambu structure explicitly in block matrix form, the inverse Green's function takes the form 
\begin{equation}
    \mathbb{\hat{D}}^{-1}_{\bf q}(z) = \begin{pmatrix}
    \mathds{1} & 0 \\
    0 & -\mathds{1} \\
    \end{pmatrix}\begin{pmatrix}
    z - \mathbb{\hat{A}}_{\bf q} & - ( \mathbb{\hat{B}}_{\bf q} - i \mathbb{\hat{C}}_{\bf q }) \\ 
     \mathbb{\hat{B}}_{\bf q} + i \mathbb{\hat{C}}_{\bf q } & z + \mathbb{\hat{A}}_{\bf q } \\ 
    \end{pmatrix}
\end{equation}
with $z$ lying in the complex plane and $\mathbb{\hat{A}},\mathbb{\hat{B}},\mathbb{\hat{C}}$ matrices in the  sublattice and bilayer subspaces.

\begin{widetext}
Provided certain conditions are satisfied by the matrices $\mathbb{\hat{A},\hat{B},\hat{C}}$, we can find the Green's function by explicitly performing the block matrix inversion procedure. 
We then find 
\begin{multline}
    \mathbb{\hat{D}}_{\bf q}(z) = \begin{pmatrix}
    \left[ z - \mathbb{\hat{A}}_{\bf q}   + ( \mathbb{\hat{B}}_{\bf q} - i \mathbb{\hat{C}}_{\bf q })(z + \hat{\mathbb{A}}_{\bf q})^{-1}(\mathbb{\hat{B}}_{\bf q} + i\mathbb{\hat{C}}_{\bf q}) \right]^{-1} & 0  \\ 
     0 & \left[ z + \mathbb{\hat{A}}_{\bf q}   + ( \mathbb{\hat{B}}_{\bf q} + i \mathbb{\hat{C}}_{\bf q })(z - \hat{\mathbb{A}}_{\bf q})^{-1}(\mathbb{\hat{B}}_{\bf q} - i\mathbb{\hat{C}}_{\bf q}) \right]^{-1} \\ 
    \end{pmatrix} \\ 
  \times   \begin{pmatrix}
    \mathds{1} & ( \mathbb{\hat{B}}_{\bf q} - i\mathbb{\hat{C}}_{\bf q})(z + \mathbb{\hat{A}}_{\bf q})^{-1} \\
    - ( \mathbb{\hat{B}}_{\bf q} +i \mathbb{\hat{C}}_{\bf q})(z - \mathbb{\hat{A}}_{\bf q})^{-1} & \mathds{1} \\
    \end{pmatrix}\begin{pmatrix}
    \mathds{1} & 0 \\
    0 & -\mathds{1} \\
    \end{pmatrix}.
\end{multline}
We can then perform all the necessary analytical continuations. 
\end{widetext}

\subsection{Photon-magnon coupling in bilayer model\label{subsec:Susceptibility}}

In this section we derive the magnon-photon coupling in the bilayer model. Throughout this section we assume the photon-electron coupling to of the scalar type $\kappa\sum_{j\in A,\bm{\delta},\ell}c_{j+\bm{\delta}\ell}^{\dagger}\hat{E}_{j,j+\bm{\delta}}c_{j\ell}$.
In Subsec.~\ref{subsec:Phonon-magnon-coupling} we derive the effective
coupling between Magnons and photons. Based on this coupling in Subsec.
\ref{subsec:Susceptibility-1} we derive the change in susceptibility
due to Magnons.

\subsubsection{Phonon-magnon coupling\label{subsec:Phonon-magnon-coupling}}

We now perform the adiabatic elimination of doublon states and derive
the effective coupling between photon and magnon degrees of freedom.
Analogously to Sec. \ref{app:bilayer} we find the scalar $H_{\text{sc}}$
and vector $H_{\text{vec}}$ terms. By expanding up to the linear
order in $\kappa$ we find the scalar coupling term:

\begin{align}
H_{\text{sc}}^{\left(2\right)} & \approx\frac{8}{U}\frac{\left(2t\right)}{\sqrt{M}}\kappa\sum_{\bm{k},\bm{q},\ell}\left(-1\right)^{\ell}\left(\hat{E}_{-\bm{q}}a_{\bm{q}-\bm{k},\ell}b_{\bm{k},\ell}+\hat{E}_{\bm{q}}a_{\bm{q}-\bm{k},\ell}^{\dagger}b_{\bm{k},\ell}^{\dagger}\right)\psi_{\bm{k}-\frac{\bm{q}}{2}}^{\left(s\right)}\nonumber \\
 & +\frac{8}{U}\frac{\left(2t\right)}{\sqrt{M}}\kappa\sum_{\bm{k},\bm{k}^{\prime},\ell}\left(-1\right)^{\ell}\hat{E}_{\bm{q}}\left(b_{\bm{k}+\bm{q},\ell}^{\dagger}b_{\bm{k},\ell}+a_{\bm{k}+\bm{q},\ell}^{\dagger}a_{\bm{k},\ell}\right)\psi_{\bm{q}}^{\left(s\right)}\label{eq:H2sc}
\end{align}
where $\psi_{\bm{k}}^{\left(s\right)}=\frac{1}{2}\left(\cos\left(k_{x}\right)+\cos\left(k_{y}\right)\right)$.
Analogously we find quadratic the vector term: 
\begin{widetext}
\begin{align*}
H_{\text{vec}}^{\left(2\right)} & =-i\frac{2\lambda}{U}\sum_{j,\delta,\ell}\left(-1\right)^{\ell}\left(\hat{c}_{j,\ell}^{\dagger}\hat{c}_{j+\bm{\delta},\ell}\hat{c}_{j+\bm{\delta},\ell}^{\dagger}\hat{E}_{j,j+\bm{\delta},\ell}\cdot\bm{\sigma}\hat{c}_{j,\ell}-\hat{c}_{j,\ell}^{\dagger}\hat{E}_{j,j+\bm{\delta},\ell}\cdot\bm{\sigma}\hat{c}_{j+\bm{\delta},\ell}\hat{c}_{j+\bm{\delta},\ell}^{\dagger}\hat{c}_{j,\ell}\right)\\
 & =\frac{8\lambda}{U}\frac{1}{\sqrt{M}}\kappa\sum_{\bm{q},\bm{q}^{\prime}}\left(-1\right)^{\ell}\left(\hat{E}_{\bm{q}^{\prime}+\bm{q}}a_{\bm{q}^{\prime},\ell}^{\dagger}b_{\bm{q},\ell}^{\dagger}+\hat{E}_{\bm{q}^{\prime}+\bm{q}}^{*}a_{\bm{q}^{\prime},\ell}b_{\bm{q},\ell}\right)\left(\sin\frac{\left(q_{x}^{\prime}-q_{x}\right)}{2}\sin\theta-\sin\frac{\left(q_{y}^{\prime}-q_{y}\right)}{2}\cos\theta\right),
\end{align*}
and linear terms:

\[
H_{\text{vec}}^{\left(1\right)}=\frac{4t\lambda i}{U}\sum_{\bm{q},\ell}\left(-1\right)^{\ell}\hat{E}_{-\bm{q}}\left(b_{\bm{q},\ell}-a_{\bm{q},\ell}+b_{-\bm{q},\ell}^{\dagger}-a_{-\bm{q},\ell}^{\dagger}\right)\left(\cos\theta\sin\frac{q_{x}}{2}+\sin\frac{q_{y}}{2}\sin\theta\right)
\]
Below we derive the susceptibility change due to the presence of magnons. 
\end{widetext}

\subsubsection{Susceptibility\label{subsec:Susceptibility-1}}

In this section we provide technical details of calculation of susceptibility
in Sec. \eqref{sec:bimagnon}. We consider the bilayer model derived
above with the anisotropy parameter $\Gamma$ and without the spin-orbit
coupling $\lambda$. With the free bilayer Hamiltonian $\hat{H}_{0}=\frac{1}{2}\sum_{\bm{k}}\Psi_{\bm{k}}^{\dagger}\mathbb{M}_{\bm{k}}\Psi_{\bm{k}}$
and the photon-magnon term $\hat{\bm{E}}_{0}\hat{V}$ with $\hat{V}=\frac{1}{2}\sum_{\bm{k}}\Psi_{\bm{k}}^{\dagger}\mathbb{V}_{\bm{k}}\Psi_{\bm{k}}$
which we get from Eq. \eqref{eq:H2sc} by setting the field momentum
to zero. The Nambu matrices are:

\[
\mathbb{M}_{\bm{k}}=\left(2J-\Gamma\psi_{k}^{\left(s\right)}\zeta_{1}\right)+\left(2J-\Gamma\right)\psi_{k}^{\left(s\right)}\tau_{1}\zeta_{1}+\frac{J_{ud}}{2}\left(1+\tau_{1}\zeta_{1}l_{1}\right)
\]

\[
\mathbb{V}_{\bm{k}}=2g^{\prime}\left(1+\psi_{\bm{k}}^{\left(s\right)}\tau_{1}\zeta_{1}\right)l_{3}
\]
The imaginary-frequency susceptibility reads:

\begin{align}
 & \chi\left(i\omega_{n}\right)\nonumber \\
 & =\int d\tau e^{i\omega_{n}\tau}\left\langle T_{\tau}\hat{V}\left(\tau\right)\hat{V}\left(0\right)\right\rangle \nonumber \\
 & =\frac{1}{\beta}\sum_{\bm{k}}\sum_{m}\text{Tr}\left\{ \mathbb{G}_{\bm{k}}\left(i\omega_{m}-i\omega_{n}\right)\mathbb{V}_{\bm{k}}\mathbb{G}_{\bm{k}}\left(i\omega_{m}\right)\mathbb{V}_{\bm{k}}\right\} ,\label{eq:chi-1}
\end{align}
where $T_{\tau}$ denotes the time-ordering and in the last line we
used Wick's theorem. We note that in Eq. \ref{eq:chi-1} the 1/4 factor
is cancelled by two identical contributions to the susceptibility
and due to the parity ($\bm{k}\rightarrow-\bm{k}$) symmetry. The
unperturbed propagator reads:

\[
\mathbb{G}_{\bm{k}}\left(i\omega_{m}\right)=\left(i\tau^{3}\omega_{m}-\mathbb{M}_{0}\right)^{-1}.
\]
In the zero-temperature limit we find:

\begin{equation}
\chi\left(i\Omega\right)=\sum_{\bm{k}}\int\frac{d\omega}{2\pi}\text{Tr}\left\{ \mathbb{G}_{\bm{k}}\left(i\omega-i\Omega\right)\mathbb{V}_{\bm{k}}\mathbb{G}_{\bm{k}}\left(i\omega\right)\mathbb{V}_{\bm{k}}\right\} .\label{eq:chi-2}
\end{equation}
In order to derive the expression for susceptibility for arbitrary
$\Gamma$ we the momentum dependence of all terms for shortness it
is convenient to parametrize the $\mathbb{M}_{\bm{k}}$ matrix as
follows:

\[
\mathbb{M}_{\bm{k}}=\alpha+\beta\zeta_{1}+\gamma\tau_{1}\zeta_{1}+\delta\tau_{1}\zeta_{1}l_{1},
\]
where $\alpha=2J+J_{\text{ud}}/2,\beta=-\Gamma\psi_{k}^{\left(s\right)},\gamma=\left(2J-\Gamma\right)\psi_{k}^{\left(s\right)},\delta=J_{\text{ud}}/2$.
In terms of these parameters the band energies are given by:

\begin{widetext}
\begin{align*}
 & \chi\left(i\Omega\right)\\
 & =4g^{\prime2}\sum_{\zeta=\pm}\int_{\bm{k}}\frac{2(\Omega_{{\bf k},+}^{\left(\zeta\right)}+\Omega_{{\bf k},-}^{\left(\zeta\right)})\left((\alpha+\beta)(\alpha+\beta-4\gamma\psi^{\left(s\right)})+(\gamma^{2}-\delta^{2})+\psi^{\left(s\right)}{}^{2}\left((\alpha+\beta)^{2}+(\gamma^{2}-\delta^{2})+\Omega_{{\bf k},+}^{\left(\zeta\right)}\Omega_{{\bf k},-}^{\left(\zeta\right)}\right)-\Omega_{{\bf k},+}^{\left(\zeta\right)}\Omega_{{\bf k},-}^{\left(\zeta\right)}\right)}{\Omega_{{\bf k},+}^{\left(\zeta\right)}\Omega_{{\bf k},-}^{\left(\zeta\right)}\left(\left(\Omega_{{\bf k},+}^{\left(\zeta\right)}+\Omega_{{\bf k},-}^{\left(\zeta\right)}\right)^{2}+\Omega^{2}\right)}
\end{align*}
\end{widetext}

\section{\label{app:cavity}Cavity Coupling}

Here we present some of the calculations used in modeling the cavity, and in particular the near-field coupling scheme we outline in the main text.
We also outline the approach we take for the calculations on cavity-induced squeezing of magnons.

\begin{figure}
    \centering
    \includegraphics[width=\linewidth]{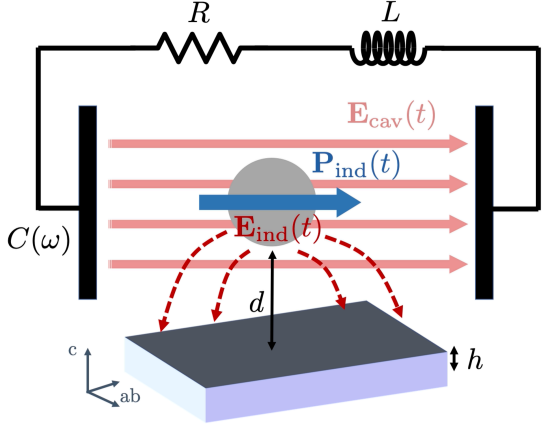}
    \caption{Schematic of the proposed near-field coupling scheme: A metal object (grey sphere) is placed inside the resonator, which we model as an RLC circuit. The cavity electric field $\mathbf{E}_{\text{cav}}(t)$ induces a dipole moment $\mathbf{P}_{\text{ind}}(t)$ in the metal object. The sample (of thickness $h)$ is placed a distance $d$ below the induced dipole, and experiences a dipole field $\mathbf{E}_{\text{ind}}(t)$ which samples momenta of order $\sim 1/d$. }
    \label{fig:near-field}
\end{figure}

\subsection{Near-Field Cavity}
We first consider the problem of a small metallic sphere with plasma response 
\begin{equation}
    \epsilon_{m}(\omega) = \epsilon_0 (1 - \Omega_{\rm pl}^2 /\omega^2 )
\end{equation}
and radius $R$.
We consider placing this within the electric field maximum of a Terahertz resonator, which we take to have a roughly homogeneous field polarized parallel to the surface of the sample.
Simple electrostatics then yields that the Terahertz field $\mathbf{E}_{\rm ext}(\omega)$ induces a dipole moment in the sphere given by 
\begin{equation}
    \mathbf{p}(\omega) = 4\pi \epsilon_0 \left[ \frac{ \epsilon_m(\omega) - \epsilon_0 }{2\epsilon_0 + \epsilon_m(\omega)}\right]R^3 \mathbf{E}_{\rm ext}(\omega).
\end{equation}
Assuming the plasma frequency is much larger than the resonator frequency, we obtain the simplified result 
\begin{equation}
\mathbf{p}(\omega) = 4\pi \epsilon_0 R^3 \mathbf{E}_{\rm ext}(\omega).
\end{equation}
We therefore see that the object serves to focus the resonator field to a subwavelength volume, producing a near-field which has the same spectral characteristics as the resonator, but with the field profile of a localized electric dipole moment. 
The induced moment produces the near-field
\begin{equation}
    \mathbf{E}_{\rm ind} = \frac{R^3}{r^3} \mathbf{E}_{\rm ext}\cdot \left[ 3 \hat{\mathbf{r}}\otimes \mathbf{\hat{r}} - 1 \right]
\end{equation}
where $\mathbf{r} = r \hat{\mathbf{r}}$ is the vector from the dipole moment to the coordinate. 

We next consider the coupling of this moment to the magnons in the sample, which we assume is placed a distance $d$ below the induced dipole, with the $ab$-plane parallel to the induced moment.
Our calculations reveal that the magnons couple to the $E_z$ component, so we Fourier transform the $E_z$ component of the induced field in order to obtain the dependence on in-plane momentum. 
We find 
\begin{equation}
    E_z(\omega,\mathbf{q}_\parallel, d) = \frac{1}{\sqrt{\rm Area}} \int d^2 \mathbf{r} e^{-i\mathbf{r}\cdot\mathbf{q}_{\parallel}} \frac{-3R^3d}{(r^2 + d^2)^\frac52} \mathbf{E}_{\rm ext}\cdot \hat{\mathbf{r}},
\end{equation}
where now we use $\mathbf{r}$ to label the coordinate with respect to the location of the dipole, projected onto the $ab$-plane a distance $d$ below.
We can integrate by parts to get 
\[
E_z(\omega,\mathbf{q}_\parallel, d) = \frac{1}{\sqrt{\rm Area}} d R^3 i \mathbf{q}_{\parallel}\cdot\mathbf{E}_{\rm ext} \int d^2 \mathbf{r} e^{-i\mathbf{r}\cdot\mathbf{q}_{\parallel}} \frac{1}{(r^2 + d^2)^\frac32} .
\]
The remaining angular integral is evaluated to be 
\[
\int d^2 \mathbf{r} e^{-i\mathbf{r}\cdot\mathbf{q}_{\parallel}} \frac{1}{(r^2 + d^2)^\frac32} = 2\pi \int_0^\infty dr \frac{ r J_0(|\mathbf{q}_{\parallel}| r) }{(r^2 + d^2 )^\frac32} ,
\]
and then we evaluate the radial integral, which yields 
\[
\int d^2 \mathbf{r} e^{-i\mathbf{r}\cdot\mathbf{q}_{\parallel}} \frac{1}{(r^2 + d^2)^\frac32} = \frac{2\pi}{d} e^{- d |\mathbf{q}_{\parallel}|} . 
\]
We therefore find the momentum space profile 
\begin{equation}
    E_z(\omega,\mathbf{q}_\parallel, d) = \frac{1}{\sqrt{\rm Area}} R^3 2\pi i \mathbf{q}_{\parallel}\cdot \mathbf{E}_{\rm ext}e^{- d |\mathbf{q}_{\parallel}|}. 
\end{equation}
The fact that this decays exponentially is related to the near-field nature of the profile.
We find that this has a maximum in momentum space at $\mathbf{q}\sim 1/d$, yielding the desired effect of sampling the coupling strength at finite momentum. 
One assumption we make is that the cavity eigenmode is not changed by the presence of the substrate, which is in general not the case. 
However, we will use this model anyways, bearing in mind the actual parameters in an experiment may be quite different, and more quantitative calculations need to be performed if device geometry is a crucial consideration. 

\subsection{Cavity Spectral Function}
We now turn our attention to the modeling of $\mathbf{E}_{\rm ext}$, which is produced by the Terahertz resonator. 
We model the system as an RLC circuit, and model the field as being generated by a parallel-plate capacitor. 
In that case, the equation of motion for the charge, in the frequency domain, is 
\begin{equation}
    \left( - I \omega^2 - i\omega R  + \frac{1}{C_{\rm eff}(\omega)} \right)Q_{\rm cav}(\omega) = 0 ,
\end{equation}
where $I$ and $R$ are the effective inductance and resistance of the circuit, and $C_{\rm eff}(\omega)$ is an effective capacitance, which includes the bare geometric capacitance $C_{0} = \epsilon_0 A_{\rm eff}/l_{\rm eff}$, as well as the dressed capacitance due to the embedded object, and its resulting coupling to the magnons below. 
We make the crude approximation that the field generated by the induced dipole is the same as the vacuum field, which obviously requires refinement in future works. 
If we consider the sample to have a thickness $h$, we find the electrostatic energy at frequency $\omega$ is determined by the functional
\begin{equation}
    U = \frac12 \epsilon_0 \mathbf{E}_{\rm cav}^2 A_{\rm eff}\ell_{\rm eff} + \frac12 \int_{d-h/2}^{d+h/2}dz \int_{\bf q} \chi_{zz}(\omega,\mathbf{q}) |\mathbf{E}_{\rm cav}\cdot \bm \epsilon_{\mathbf{q}}(z) |^2,
\end{equation}
with 
\begin{equation}
    \bm \epsilon_{\bf q}(z) = 2\pi R^3 i \mathbf{q} e^{-zq}. 
\end{equation}
We identify the total capacitance as $C_{\rm eff} = \frac{1}{\ell_{\rm eff}^2} \delta^2 U /\delta^2 E $, which gives 
\begin{equation}
C(\omega) =  \epsilon_0 \frac{A_{\rm eff}}{\ell_{\rm eff} }+ \frac{4\pi^2 h R^6}{\ell_{\rm eff}^2}  \int_{\bf q} \chi_{zz}(\omega,\mathbf{q}) |\mathbf{\hat{e}}_{\rm cav}\cdot \mathbf{q}|^2 \frac{\sinh hq e^{-2d q}}{hq}.
\end{equation}
The second contribution is regarded as the capacitance between the resonator and substrate, or alternatively, is the substrate's contribution to the energetic cost of the induced dipole.
Therefore, the capacative coupling to the magnons is given by 
\begin{equation}
C_{\rm mag}(\omega) =   \frac{4\pi^2 h R^6}{\ell_{\rm eff}^2}  \int_{\bf q} \chi_{zz}(\omega,\mathbf{q}) |\mathbf{\hat{e}}_{\rm cav}\cdot \mathbf{q}|^2 \frac{\sinh hq e^{-2d q}}{hq}.
\end{equation}
We therefore end up with the dressed cavity spectral function (up to an overall constant) of 
\begin{equation}
    A_{\rm cav}(\omega) = -\frac{1}{\pi} \Im \frac{1}{\omega^2 + i\gamma \omega - \Omega_0^2(1+ K_{\rm mag}(\omega) )^{-1}}
\end{equation}
with the usual relations $\gamma = R/I, \Omega_{0}^2 = 1/(IC)$ and the dimensionless coupling function 
\begin{equation}
K_{\rm mag}(\omega) = \frac{4\pi^2 h R^6}{\ell_{\rm eff} A_{\rm eff} a^4 }  \int_{\bf q} \frac{ \chi_{zz}(\omega,\mathbf{q}) }{\epsilon_0}|\mathbf{\hat{e}}_{\rm cav}\cdot \mathbf{q}|^2 \frac{\sinh \tfrac{h}{a}|\mathbf{q}| e^{-2\tfrac{d}{a}|\mathbf{q}|}}{\tfrac{h}{a}|\mathbf{q}|}.
\end{equation}
In the above expression we have replaced the lattice unit of $a\sim \SI{4}{\angstrom}$, originating from the momentum integral, such that now $\mathbf{q}$ is actually dimensionless.

We regard the geometric factors as somewhat phenomenological, in that they are ``flexible" with respect to our model, and should be determined from actual device parameters or simulations.
For a rough estimate, we take $d/a \sim R/a \sim 100$, $h/a \sim 10$, and  $ \ell_{\rm eff} A_{\rm eff} /a^3 \sim 10^6$.
Note this corresponds to $ d \sim R \sim 40\si{\nm}$, $h\sim 4\si{\nm}$, and an effective volume of $\ell_{\rm eff}A_{\rm eff} \sim 40 \si{\nm} \times 40 \si{\nm} \times 40 \si{\nm}$.
If we approximate $\chi$ by the $F| \mathbf{q}\cdot \mathbf{N}|^2/ (\Omega^2 - \omega^2)$ and consider the case of $\mathbf{N} = \mathbf{\hat{e}}_{\rm cav} = \mathbf{e}_x$ we get a rough estimate for $K$ of 
\[
K_{\rm mag} \sim \frac{3 \cdot 5!}{8\cdot 2^6 } \frac{h (R/d)^6 a^2}{\ell_{\rm eff} A_{\rm eff}} \frac{F }{\Omega_{\rm sw}^2 - \omega^2}
\]
where $F\sim 6.2 \times 10^4 \si{\meV}^2$ is the oscillator strength estimated in the main text. 
The numerical prefactor is of order one, and so the main suppression is due to the disconnect between the capacitor size and the lattice scale, with $h a^2/ \ell_{\rm eff} A_{\rm eff} \sim 10^{-5}$.

Bearing this in mind by approximating 
\begin{equation}
K_{\rm mag}(\omega) =  \alpha^2 \frac{F}{\Omega_{\rm sw}^2 - \omega^2 } .
\end{equation}
where $\alpha^2 \sim 10^{-5} $ is the effective coupling constant, we obtain the spectral function 
\begin{equation}
    A_{\rm cav}(\omega) = -\frac{1}{\pi} \Im \frac{1}{\omega^2 + i\gamma \omega - \Omega_0^2\left(1+ \frac{\alpha^2 F}{\Omega_{\rm sw}^2 - (\omega + i0^+)^2 } \right)^{-1}}.
\end{equation}
Recall that typical values for $\Omega_{\rm sw} $ and $F$ are $\Omega_{\rm sw} \sim 1 \si{\THz}$ and $F \sim 3.8 \times 10^3 (\si{\THz})^2 $.

This can reveal a clear avoided crossing in frequency space as the cavity resonance is tuned through the spin-wave resonance. 
This doesn't take dispersion of the spin-wave into account, for instance, and this could possibly spoil the effect since dispersion of the resonance will act as a sort of inhomogeneous broadening.
Depending on the characteristic momentum of the magnon sampled by the geometry, this may be important. 

If we are going to be more careful, we see that the correction due to the dispersion of the mode is approximated as 
\begin{equation}
    \Omega_{\bf q}^2 = \Omega_0^2 + c^2 \mathbf{q}^2 /a^2
\end{equation}
with dimensionless momentum $\mathbf{q}$ and $c= \sqrt{2}J_0$.
Then, we find for the response function $K$, in the limit of $h \ll d$, and including the angular dependence as well 
\begin{equation}
    K_{\rm mag}(\omega) = 5! \pi \frac{ h a^2}{V_{\rm eff} } \left(\frac{R}{2d}\right)^6 \left[ \frac34(\mathbf{\hat{N}\cdot \hat{E}})^2 + \frac14 (\mathbf{\hat{N}\times\hat{E}})^2 \right] \frac{F}{\Omega_0^2} g(z, \kappa) 
\end{equation}
with the function
\begin{equation}
    g(z,\kappa) = \frac{1}{5!}\int_0^{\infty} dx x^5 e^{-x} \frac{1}{1+\kappa^2 x^2 - (z+i0^+)^2 }
\end{equation}
where we have introduced 
\begin{subequations}
\begin{align}
& z = (\omega + i0^+)/\Omega_0 \\
& \kappa = \frac{c}{2d \Omega_0} .
\end{align}
\end{subequations}
We have also used $\mathbf{\hat{N},\hat{E}}$ to indicate the unit vectors along the N\'eel vector and in-plane cavity polarization, respectively. 
The previous calculation holds in the limit of $\kappa^2 \sim \frac{a^2}{d^2}\frac{J_0}{\Gamma} \to 0$, in which case the function $g$ reduces to the clear resonance at $z = 1$.
On the other hand, for $\kappa \gtrsim 1$, we cannot discount the dispersion effects, and this will quickly destroy the resonance due to the inhomogeneous broadening of the dispersing magnon.
While the ratio $a^2/d^2$ is clearly quite small (and part of the goal is to enlarge this parameter), the ratio $J_0/\Gamma \sim 10^4$ is also quite large, reflecting the rapid onset of dispersion.
We have inadvertently chosen $d/a\sim 10^{2}$ such that $\kappa \sim 1$. 
If we decrease $\kappa$ to a value of order $.1$, we can probably resolve the resonance.
This means increasing $d/a$ to be around $d/a\sim 300$, which will in turn decrease the overall size of the coupling constant by $1/3^6$.
This would kill the effect, unless we concomitantly scale $R$ such that $R/d$ remains constant. 
This means $R \sim d \sim 300 a\sim 1200 \si{\angstrom}$.

\end{document}